\documentclass[a4paper,UKenglish,numberwithinsect]{lipics-v2021}
\PassOptionsToPackage{nocompress}{cite}
\usepackage{amsmath,amssymb,amsthm,mathtools}
\usepackage{thm-restate}
\usepackage{microtype}
\usepackage{enumitem}
\usepackage{xcolor}
\usepackage{tikz}
\usepackage{cite}
\usetikzlibrary{arrows.meta,patterns}
\tikzset{
  qtcell/.style={draw=blue!70!black,thick},
  qtrange/.style={draw=black!65,dashed},
  qtcut/.style={draw=orange!80!black,densely dashed,thick},
  qthighlight/.style={color=orange!80!black},
  qtnode/.style={draw,rounded corners=2pt,fill=white,inner sep=4pt},
  qtedge/.style={-{Stealth[length=4pt]},semithick}
}

\nolinenumbers
\hideLIPIcs

\title{Dynamic Quadtrees and the Real-RAM}

\author{Sarita de Berg}{IT University of Copenhagen, Denmark}{debe@itu.dk}{0000-0001-5555-966X}{}

 \author{Jacobus Conradi}{University of Copenhagen, Denmark}{jaco@di.ku.dk}{}{}

 \author{Ivor van der Hoog}{IT University of Copenhagen, Denmark}{ivva@itu.dk}{https://orcid.org/0009-0006-2624-0231}{}

 \author{Eva Rotenberg}{IT University of Copenhagen, Denmark}{erot@itu.dk}{0000-0001-5853-7909 }{}{}

\funding{This work was supported by the VILLUM Foundation grant (VIL37507) ``Efficient Recomputations for Changeful Problems'' and and Danmarks Frie Forskningsfond via the grant ``Dynamic Graphs: Distributed and Geometry'' with grant-ID 10.46540/4251-00004B and by the Carlsberg Foundation, grant CF24-1929.
}

\authorrunning{Sarita de Berg, Jacobus Conradi, Ivor van der Hoog, Eva Rotenberg}

\Copyright{Sarita de Berg, Jacobus Conradi, Ivor van der Hoog, Eva Rotenberg}

\hideLIPIcs

\theoremstyle{definition}
\newtheorem{invariant}{Invariant}

\newcommand{\fixedEpsilon}[1][1]{\ensuremath{\frac{#1}{12}}}
\newcommand{\fixedDelta}[1][1]{\ensuremath{\frac{#1}{56}}}
\newcommand{\R}{\mathbb{R}}
\newcommand{\diam}{\operatorname{diam}}
\newcommand{\dist}{\operatorname{dist}}

\begin{document}
\maketitle

\begin{abstract}
Dynamic quadtrees are a fundamental tool in computational geometry.
They are used to store sets of geometric objects (typically, points or disks) and
underlie dynamic algorithms for proximity queries,
stabbing queries,
well-separated pair decompositions (WSPDs),
geometric spanners,
connectivity in geometric intersection graphs,
geometric independent sets,
and more.
These applications often use quadtree compression to keep the tree small.
However, we survey a wide range of constructions and show that none of them establishes how to maintain a compressed quadtree for arbitrary real-valued point sets on a Real-RAM in sublinear worst-case time per insertion, even in the insertion-only setting.
We define and maintain a dynamic quadtree that can support the operations used in \emph{some} of them; establishing their results for arbitrary real-valued inputs (at a slowdown).
For the remaining applications, we do not establish the corresponding bounds on a Real-RAM.
In particular, for arbitrary real-valued inputs, we are unaware of any construction that establishes sublinear update time for an explicit well-separated-pair decomposition on a Real-RAM.
\end{abstract}

\subparagraph{AI statement.}
The core ideas and research direction in this paper are entirely due to the authors.
We used ChatGPT Astra 6.2 and Claude Opus 5.5 to assist in proving results, searching the literature, proofreading and editing. We also used ChatGPT, Claude, and Google DeepReview to check the proofs for errors, and all three certified the results as correct. The authors have spent considerable effort in proving theorems, proofreading, and editing, and we take full responsibility for the correctness of the paper.

\newpage
\section{Introduction}

Dynamic quadtrees are a fundamental tool in computational geometry.
They are used to store sets of geometric objects (typically points or disks) and
underlie dynamic algorithms for proximity queries%
~\cite{Eppstein2008Skip,Chan2006Minimalist,MountPark2010,ParkMount2012,Park2013Thesis,VenkatMount2014},
stabbing queries~\cite{KhramtcovaLoffler2017,HoogLoffler2018Stabbing},
well-separated pair decompositions
~\cite{FischerHarPeled2005},
geometric spanners%
~\cite{BhoreToth2021Online,Bhore2022Online,deBerg2026DiskSpannerFull},
connectivity in geometric intersection graphs%
~\cite{Baumann2024Disks,Hoog2024SquareConnectivity,deBerg2026DiskSpannerFull},
geometric independent sets%
~\cite{Bhore2020MIS,Henzinger2020MIS,BhoreChan2025},
and more%
~\cite{AcarHudson2008,Acar2013WellSpaced}.
As our first contribution, we survey a wide collection of dynamic algorithms and how they use quadtrees. We show that none of the surveyed constructions establishes how to maintain a compressed quadtree for arbitrary real-valued point sets on a Real-RAM in sublinear worst-case time per insertion.
As our second contribution, we present a dynamic quadtree that can support the operations used in \emph{some} of the surveyed papers, establishing their results for arbitrary real-valued inputs (at a slowdown).

\subparagraph{The Real-RAM.}
We use the definition of the Real-RAM model of computation from the textbook by Preparata and
Shamos~\cite{preparataShamos1985}. Each memory cell stores a real number or
an integer. Reading, writing, basic arithmetic operations, and comparisons
take constant time. In the Word-RAM model of computation, one can use the output of arithmetic computations to index memory.
On a Real-RAM,  an arithmetic
operation involving a real number returns a real number which cannot
be used as an address. The model does not provide rounding, integer division,
or bit operations on real values.

\begin{figure}[b]
    \centering
    \includegraphics{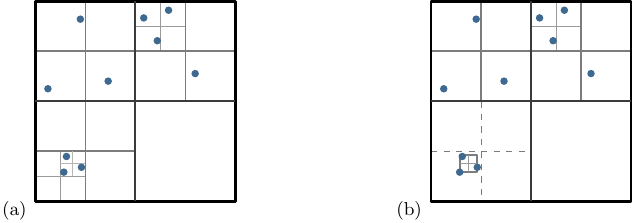}
    \caption{(a) An  uncompressed and (b) compression using the smallest bounding square.}
    \label{fig:intro}
\end{figure}

\subparagraph{Quadtrees.}
A \emph{split} of a square divides it along its midlines
into four squares called \emph{quadrants}.
In its purest form, a quadtree recursively splits a square to create a tree, see Figure~\ref{fig:intro}(a).
To maintain linear space, quadtrees require \emph{compression} which we abstract as follows:

\begin{definition}
A \emph{compressed quadtree} over $B$ is a rooted tree
whose nodes are squares called \emph{cells}.
The root cell is $B$.
Each node with cell $C$ is of one of the following types:
\begin{itemize}[nolistsep]
    \item A \emph{leaf} has no children.

    \item A \emph{regular node} has four children, whose cells are the
    four quadrants of $C$.
    \item A \emph{compressed node} has only one child $C'$. We call the edge to $C'$ a \emph{compressed link}.
\end{itemize}
For any planar point set $P \subset B$ of $n$ points, a compressed quadtree \emph{stores} $P$ if each point in $P$ is stored in exactly one leaf that contains it and contains no other points of $P$.
\end{definition}

We represent quadtrees explicitly: each node stores its cell’s coordinates, together with pointers to its children.
For a cell $C$ with a compressed compressed link to $C'$, we call $C'$ \emph{canonical} if it can be obtained from $C$ by regular splits.
On a Word-RAM, a maximal path from a cell $C$ to a descendant 
$C'$ that contains the same stored points can be replaced by a compressed link to a canonical cell.
 If the path
has $k$ edges, then $|C'|=2^{-k}|C|$.
Word operations can compute this canonical cell $C'$ in constant time. On the other hand, ordinary real arithmetic does not supply the same
constant-time operation.
For any dynamic compressed quadtree, is it crucial that the dynamic algorithm specifies  \emph{how} it chooses, computes, and maintains
that bounding square.
The \emph{Handbook of Discrete and Computational Geometry}%
~\cite[Section~32.2]{MitchellMulzer2017Proximity} discusses this problem, referring to~\cite[Appendix~A]{BuchinLofflerMorinMulzer2011}, who instead define $C'$ as a bounding square of the remaining points.
This paper surveys six approaches in the existing literature to compute compressed cells. Appendix~\ref{app:literature}
gives the definitions, computational assumptions and relevant arguments
for each paper:
\begin{enumerate}
    \item\label{fin_coor} \emph{Finite-coordinate implementations.}
    Papers that, throughout, use Word-RAM operations%
    ~\cite{Chan2002RAM,Chan2006Minimalist,VenkatMount2014,ChanSkrepetos2015,ChanYu2025}.
    These are restricted to the Word-RAM model of computation.
    \item\label{unbounded_time} \emph{Unbouded time and space.}
    Papers that  explicitly forgo compression, explicitly spending time and space linear in the number of splits needed to create all nodes~\cite{BennettYap2017,Henzinger2020MIS,Agarwal2021Similarity,AcarHudson2008,Acar2013WellSpaced}.
    For arbitrary real-valued input, the number of splits needed is unbounded.

    \item\label{restricted_construction} \emph{Restricted constructions.}
    Papers that work on real-valued input, but then restrict to a fixed range of object sizes%
    ~\cite{Baumann2024Disks,deBerg2026DiskSpannerFull, Kaplan2020Voronoi}, or restrict compressed links~\cite[Section~7.7]{Hoog2022Thesis}
    (which corrects~\cite{Hoog2018Smooth}).
    Again, these use unbounded time under arbitrary real-valued inputs.

    \item\label{additional_operations} \emph{Additional real-coordinate operations.}
    Papers that work on real-valued input, and additionally assume floor or coordinate-bit operations to compute canonical cells~\cite{AluruSevilgen1999,KisfaludiBak2026Steiner,Bhore2020MIS,Eppstein2008Skip,MountPark2010,ParkMount2012,Park2013Thesis,GudmundssonHuangWong2023,chan2019Locality,GaoHarPeled2025LSO}.
    Appendix~\ref{app:literature-strong} shows that adding floor allows a Real-RAM to solve all NP problems in polynomial time. Other augmentations even imply that P=PSPACE.

    \item\label{unspecified_op} \emph{Unspecified operations.}
    Papers that leave the construction of the cell $C'$ unspecified    ~\cite{FischerHarPeled2005,Hoog2024SquareConnectivity,BhoreChan2025,BhoreToth2021Online,Bhore2022Online,HoogLoffler2018Stabbing, KhramtcovaLoffler2017}. As written, their algorithmic description is therefore incomplete.
    
    \item\label{smallest_bounding_square} \emph{Smallest bounding squares.}
    Papers that choose $C'$ as a smallest bounding square
    of the points in $P \cap C$~\cite{LofflerSimonsStrashFull}, see Figure~\ref{fig:intro}(b).
    We define this as a \emph{strict quadtree}
    (Definition~\ref{def:strict}).
\end{enumerate}

For papers in category  
2 and 3 the usage of dynamic quadtrees is unproblemmatic but worst-case inefficient. 
For categories 4, 5, and 6, the proofs as written may need revision for the results to hold.
We emphasize that many of these results may very well be true, they just need a different proof that uses compressed dynamic quadtrees differently or not at all. 

\subparagraph{Contribution.}
Our first contribution is the above survey, which identifies which dynamic implementations require finite coordinates, are required to assume bounded object spread, assume overly strong real-coordinate
operations, or leave required operations unsupported.
Our second contribution is that for strict quadtrees, we prove that a sequence of $n$ insertions can force
$\Omega(n^2)$ total work, regardless of the dynamic maintenance algorithm (Theorem~\ref{thm:strict-lower-bound}).

In an attempt to remedy the above issues, we isolate the \emph{quadtree packing property}
(Definition~\ref{def:packing}). This property is critical in  dynamic
proximity searching, well-separated pair decompositions, and disk
intersection graph  connectivity~\cite{Eppstein2008Skip,VenkatMount2014,FischerHarPeled2005,Acar2013WellSpaced,Baumann2024Disks,Hoog2024SquareConnectivity,deBerg2026DiskSpannerFull}.
We then show how to dynamically maintain a compressed quadtree with this property
using $O(n)$ space and $O(\log n)$ worst-case insertion and deletion time (Theorem~\ref{thm:main}). 
We define our quadtree in Section~\ref{sec:our_quadtree}. In the remainder, we show that our tree recovers \emph{some} prior results.


We recover the classical dynamic planar point storage~\cite{AluruSevilgen1999} in Section~\ref{app:extension} (Theorem~\ref{thm:extensions}). We can store points in $\mathbb{R}^d$ in a linear size tree and the node of a point can be found in logarithmic time and insertions and deletion require also logarithmic time (for fixed $d$).

We also recover a data structure for dynamic approximate nearest-neighbour queries in Appendix~\ref{app:ann-recovery} (Theorem~\ref{thm:ann-recovery}). In this problem, the goal is to maintain a data structure which returns for a query point $q$, a point that is at most a factor $\varepsilon$ further from $q$ compared to its nearest neighbour. Compared to \cite{Chan2006Minimalist,VenkatMount2014}, our result incurs extra $\log n$ and $\varepsilon$ factors.

We recover only \emph{implicit} well-separated pair decompositions (WSPD) in Appendix~\ref{app:wspd}: rather than maintaining an explicit list of all pairs, we can maintain a tree where at all times, the classical tree-to-WSPD conversion algorithm yields an $s$-WSPD in $O(s^2 n)$ time (Theorem~\ref{thm:wspd}).
The dynamic WSPD algorithm of Fischer and Har-Peled~\cite{FischerHarPeled2005} assumes that, once the position of a new branching node is known, its canonical cell can be computed in constant time. They leave open how to compute this cell on a Real-RAM. 
In Appendix~\ref{app:callahan}, we
identify errors and gaps in the dynamic WSPD approach
of~\cite{CallahanKosaraju1995Dynamic,Callahan1995Thesis}.
Thus, whether one can update an explicit $s$-WSPD 
in sublinear worst-case 
time on a Real-RAM remains
open and we conjecture that it can not.
Appendix~\ref{app:literature} discusses how each paper makes use of compressed quadtrees and whether their results are or aren't partially recovered by our tree.

\subparagraph{Organisation and preliminaries.} In Section~\ref{sec:strict-lower-bound} we prove an amortised linear-time lower bound for strict dynamic quadtrees and in Section~\ref{sec:our_quadtree} present our new dynamic quadtree.
In the main body, we consider a quadtree over a fixed square $B \subseteq \mathbb{R}^2$ that stores a dynamic point set $P$ contained in $B$. 
In Section~\ref{app:extension}, we generalise to not having a fixed bounding box and generalise to $\mathbb{R}^d$.
To decide which square owns a point on a shared boundary,
we use the following convention: 
Except for when we discuss strict quadtrees, all rectangles in this paper are axis-aligned, closed along their bottom and
left boundaries, and open along their top and right boundaries.
We denote the length of its longest side by $|R|$.

\section{A lower bound for strict quadtrees}
\label{sec:strict-lower-bound}

We show a linear lower bound for the amortised update time of a dynamic strict quadtree:

\begin{definition}[Strict quadtree]
\label{def:strict}
A \emph{strict quadtree} is a compressed quadtree where we require for each  compressed node $C$ with child $C'$ that: (i)
$C'\subseteq C$, and (ii) $|C'|\leq |C|/4$, and (iii) $C'$ is a smallest
closed square containing the points stored in its subtree.
\end{definition}

For strict quadtrees, we need to assume that the cells are closed on all sides. Otherwise, the smallest square containing the points is not well-defined. In this case, to make sure that for each point there is still a single root-to-leaf path whose cells contain the point, any point on a vertical (respectively horizontal)
midline for a regular split is assigned to the quadrant above (respectively to the right of) the point.

\begin{theorem}
\label{thm:strict-lower-bound}
For every deterministic algorithm to maintain a dynamic strict quadtree, 
there is a sequence of $n$ point insertions that requires $\Omega(n^2)$
total update time.
\end{theorem}

\begin{proof}
We prove the lower bound in one dimension, where quadtree cells are intervals
and a regular split bisects its interval. The construction also holds in higher dimensions.

Fix some integer $m$ and set the root $B$ of the strict quadtree as $[0,1]$. We provide a sequence of $n \in O(m)$ insertions that gives the $\Omega(m^2)$ lower bound on the total update time. We first provide a sequence of point insertions that forces many specific splits, and then provide the remaining point insertions and show that each requires $\Theta(m)$ work.

We insert a point at $0$, and all other points in the interior of the interval $I=[1-2^{-m^2},1]$.
Set $x=1-7|I|/40$.
We first insert $x$, the two points
$a=x-|I|/8$ and $b=x+|I|/8$, and the $2m$ points $p_i$ and $p_i'$ for $i\in\{0,\ldots,m-1\}$ (see Figure~\ref{fig:lower-bound-points}):
\[
  \qquad p_i := x+r_i, \qquad p_i' := x-r_i,\ \qquad \text{ where }
 \qquad r_i=\frac{|I|}{12}\left(\frac34\right)^i.
\]
We thus insert $2m+3$ distinct points, whose smallest bounding interval is
$[a,b]$.

\begin{figure}
\centering
\includegraphics{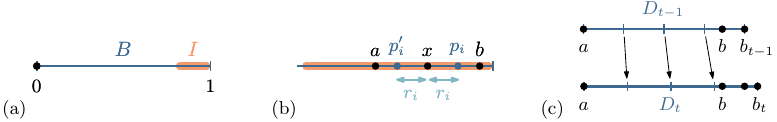}
\caption{
(a) The root $B$ and the interval $I$. All inserted points except $0$ lie in $I$.
(b) Pairs $p_i',p_i$ are at distance $r_i$ from $x$ and approach $x$.
(c) Inserting the point $p_t$ forces replacing $D_{t-1}$ by $D_t$. This moves its midpoint
and changes the sizes and coordinates at every depth of its regular split.}
\label{fig:lower-bound-points}
\end{figure}

\begin{claim}\label{claim:regular_splits}
    For any interval $J=[u,v]\subsetneq I$ containing all points in~$I$ inserted so far and any subtree rooted at~$J$ of a strict quadtree storing them (and possibly some more points), the path from $J$ to the leaf containing $x$ begins with $\Theta(m)$ regular splits.
\end{claim}
\begin{proof}[Proof of Claim~\ref{claim:regular_splits}]

Consider a cell $C$ on this path with $|C|\geq 5r_{m-1}/2$,
and suppose all splits from $J$ up to $C$ are regular. We next show that $C$ must also be a regular node.

If $|C|\geq |I|/3$, we argue that $C$ contains both $a$ and $b$. Since $v-x\leq 7|I|/40<|I|/3$, $C$ is the rightmost cell of this size in $J$.
Furthermore, the left endpoint of the interval $C$ is at most $1-|I|/3<a$, and thereby $C$ contains both points $a$ and $b$.
The interval spanned by the points in $C$  thus has size at least $b-a = |I|/4>|C|/4$,
so $C$ must be a regular node.

If instead $|C|<|I|/3$, we argue that there is a point $p_i$ or $p_i'$ in $C$ with $r_i > |C|/4$.
Because
$4r_{i+1}=3r_i>5r_i/2$ and $4r_0=|I|/3$, the ranges $[5r_i/2,4r_i)$ overlap.
Hence there is some $i\in\{0,\ldots,m-1\}$ for which
$|C|/4<r_i\leq 2|C|/5$.
One endpoint of $C$ lies at distance at least $|C|/2$ from $x$,
so the point $p_i$ or $p_i'$, which is at distance $r_i$ from $x$, towards this endpoint lies inside $C$.
Thus, $C$ contains two points ($x$ and $p_i$ or $p_i'$) whose distance is greater than $|C|/4$.
We again conclude that $C$ is a regular node.

By induction, the path repeatedly halves its cell from size
$|J|\geq b-a = |I|/4$ until falls below $5r_{m-1}/2$.
Since $r_{m-1}=(|I|/12)(3/4)^{m-1}$, we find $\Theta(m)$
consecutive regular splits.
\end{proof}

\subparagraph*{The remaining insertions.}
We next insert $m$ points $b_t$ with $
 b<b_1<\cdots<b_m<x+\frac{3|I|}{20}$. 
After inserting $b_t$, the smallest bounding interval of the points in $I$ is
$D_t=[a,b_t]$, with $|I|/4<|D_t|<11|I|/40$.
We choose each $b_t$ so that no previously stored cell has size $|D_t|/2^j$ for any $j\geq0$.
Since each stored cell of size $\lambda$ forbids only the countably many
values $a+2^j\lambda$ and there are uncountably many
reals in $(b_{t-1},x+3|I|/20)$, such a choice for $b_t$ exists.

\subparagraph*{The work after each insertion.}
After inserting $b_t$, all points except $0$ lie in $D_t=[a,b_t]$.
If the path towards $I$ has no compressed link before reaching $I$,
the tree already contains $m^2$ regular nodes and the lower bound follows.
Otherwise, the first compressed link contains points except $0$, and property~(iii) makes its child $D_t$.
By Claim~\ref{claim:regular_splits} with $J=D_t$, we obtain
$\Theta(m)$ regular cells of sizes $|D_t|,|D_t|/2,|D_t|/4,\ldots$.
By our choice of $b_t$, none of these sizes was previously stored.
Thus, each of the $m$ remaining insertions requires $\Omega(m)$ cell
creations or coordinate changes, see Figure~\ref{fig:lower-bound-points}(c). Since nodes explicitly store their cell
coordinates, this gives a total update time of $\Omega(m^2)$.
With $n=3m+4$ inserted points, this is $\Omega(n^2)$.
\end{proof}

\subsection{Can we not simply fix this?}

Our lower bound abuses property~(iii): a compressed child \emph{must} be
a smallest bounding cell of its stored points.
Inserting $b_t$ then changes this smallest bounding cell, and all the regular cells below it.
We review four simple fixes, to show that these each have their own problems.

\subparagraph*{(1) Upgrowing $\mathbf{B_{t-1}}$.}
One may try to preserve the old component by adding a parent node for $B_{t-1}$ that contains $b_t$. The parent node is then twice as large, an to include $b_t>b_{t-1}$, it must thus be 
$[a,2b_{t-1}-a]$.
However, this interval extends beyond the boundary of $B$.

\subparagraph*{(2) Aligning the smaller component.}
Another approach would be to merge the two components when an update makes the child of a compressed link larger than one quarter of its parent in side length. We would then align the component with fewer nodes to the other componen, spending time proportional to its number of nodes (as suggested in~\cite[Appendix~C.1]{LofflerSimonsStrashFull})
Here, the smaller component is the singleton $b_t$.
Keeping the existing grid requires \emph{upgrowing} $B_{t-1}$, meaning we are essentially back in case (1).

\subparagraph*{(3) Coinciding with the boundary.}
One may be tempted to snap a compressed child $\alpha$ to the
boundary of its parent $\beta$ (i.e., choosing $[a,1]$). 
However, a later insertion may place a new node $\gamma$ that is a direct ancestor of $\beta$.
If $\beta$ lies close to the boundary of $\gamma$, we must snap $\beta$ to $\gamma$. But, then we must re-snap $\alpha$ (and all its descendents) to $\beta$.
If each such insertion forces us to re-snap a linear-size descendant
component, repeating  takes
$\Omega(n^2)$ total time.

\subparagraph*{(4) Allowing cells to stick out.}
Finally, one could also drop the requirement that compressed cells should be contained in their parents. This is non-trivial, as it breaks several quadtree properties including the property that each point lies in a unique root-to-leaf path.

\section{A new aligned quadtree}\label{sec:our_quadtree}

\begin{figure}[h]
\centering
\includegraphics[]{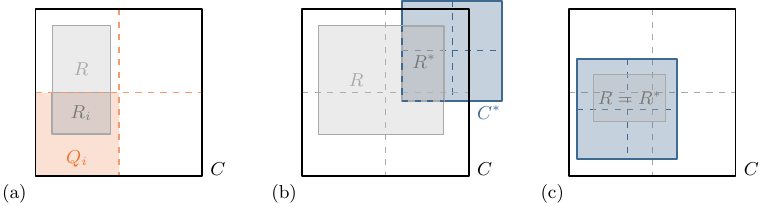}
\caption{
(a) A regular node with two children that have an empty range and are therefore leaves.
(b) A compressed node with a child that is not contained in its parent.
(c) A compressed node with a blue child $R^*$ that has the exact same range as its parent, but is not aligned with it. }
\label{fig:cell-range-overview}
\end{figure}

We design a new compressed quadtree that we call the \emph{aligned quadtree}. 
We show that we can dynamically maintain an aligned quadtree with a useful packing property. 
In contrast to old quadtree definitions, our quadtree nodes consist of a cell $C$, which is a square, together with a \emph{range} $R$, which is a rectangle contained in $C$.
The key idea is that a cell $C$ only stores points in its range. Formally, quadtree nodes are defined as follows, see Figure~\ref{fig:cell-range-overview}.

\begin{definition}
\label{def:nodes}
Each node of an \emph{aligned quadtree} is of one of the following types:
\begin{itemize}[nolistsep]
 \item 
 A \emph{leaf} $(C,R)$ has no children and stores at most one point, which lies in $R$.

 \item 
 A \emph{regular node} $(C,R)$ has four children $(Q_i,R\cap Q_i)$, one for each quadrant $Q_i$ of $C$.
 A child with an empty range is a leaf.

 \item
 A \emph{compressed node} $(C,R)$ has one child $(C^*,R\cap C^*)$ with $R \cap C^*$ non-empty which is not a leaf. We call the edge to $(C^*,R\cap C^*)$ a \emph{compressed link}.
\end{itemize}
\end{definition}

\begin{definition}[Aligned quadtree]
\label{def:tree}
An \emph{aligned quadtree} over $B$ is a tree
with root $(B,B)$ and nodes as in Definition~\ref{def:nodes}.
It \emph{stores} a set $P\subseteq B$ of $n$ points if each point in $P$
is stored in exactly one leaf whose range contains it
and stores no other points.
We denote such an aligned quadtree by $T_B(P)$.
A node is \emph{occupied} if its subtree contains a stored point.
\end{definition}

\subparagraph{Point location in a quadtree.} For any query point $q\in B$, there is either a unique leaf whose range contains $q$, or a unique compressed node $(C, R)$ with child $(C^*, R^*)$ where $q \in R \setminus R^*$.
We can find such a node top-down as follows. When we encounter a regular node, we follow the unique child whose range contains $q$.
At a compressed node, we follow its child if the child cell contains $q$, otherwise, we report the compressed node.
At a leaf, we report the leaf. 

However, because a quadtree does not have logarithmic height, searching from the root downwards may take too long. 
We can instead use an edge-oracle tree to find the stopping node in $O(\log n)$ time~\cite[Sections~2.1--2.2]{LofflerSimonsStrashFull}. 
For completeness, Appendix~\ref{app:edge-oracles} shows how top trees~\cite{Alstrup2005TopTrees} implement edge-oracle trees with
$O(\log n)$ query and update time.

\begin{restatable}[Appendix~\ref{app:edge-oracles}]{theorem}{edgeoracletheorem}
\label{thm:edge-oracles}
Our quadtrees admit a dynamic edge-oracle tree with $O(n)$ space
and $O(\log n)$ query and update time.
\end{restatable}

\subsection{Properties of our quadtree}

Many results using quadtrees rely on some packing property of the quadtree. We formalise this property in the following definition.

\begin{figure}[b]
\centering
\includegraphics[]{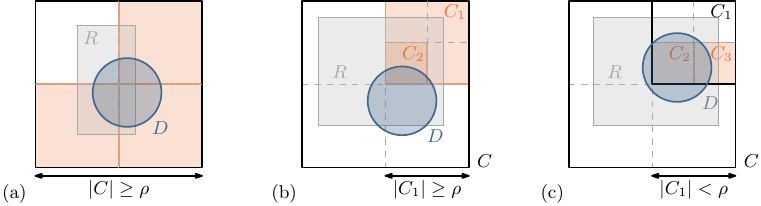}
\caption{Illustration of the packing property. Orange cells are candidates for $\Pi$.
(a) The three ranges intersect $D$ and their lowest common ancestor $C$ has $|C|\geq\rho$, so these three nodes may be in $\Pi$ together.
(b) $C_2$ is a descendant of $C_1$, and thus they may not be in $\Pi$ together.
(c) $C_2$ and $C_3$ can not both be in $\Pi$ because their parent $C_1$ has side length
less than $\rho$.}
\label{fig:packing-definition}
\end{figure}

\begin{definition}[Quadtree packing property]
\label{def:packing}
An aligned quadtree has the \emph{quadtree packing property}
if there is a constant $M$ with the following guarantee.
Let $D$ be any disk of radius $r>0$, and let $\rho>0$.
Let $\Pi$ be a set of quadtree nodes whose ranges intersect or
touch~$D$ such that (1)  no node in $\Pi$ is an ancestor of another, and  (2) 
the cell of the lowest common ancestor of any two nodes in $\Pi$
has side length at least $\rho$.
Then $\Pi$ contains at most $ M\left(1+\frac r\rho\right)^2$ nodes. 
\end{definition}
Figure~\ref{fig:packing-definition} illustrates which cells may contained together in $\Pi$.

For arbitrary compressed quadtrees, we can define the same property as above by simply replacing ``ranges'' by ``cells''. A simple packing argument shows that a strict quadtree has the quadtree packing property for $M=4\pi$.
However, as shown in Section~\ref{sec:strict-lower-bound} it can not be efficiently dynamically maintained. In Section~\ref{sec:insertions} and Section~\ref{sec:deletions} we show that points can be inserted into and deleted from an aligned quadtree, and it can thus be maintained efficiently. In Section~\ref{sec:quadtree_packing} we then prove that our aligned quadtree also has the packing property.

\begin{figure}
\centering
\includegraphics[]{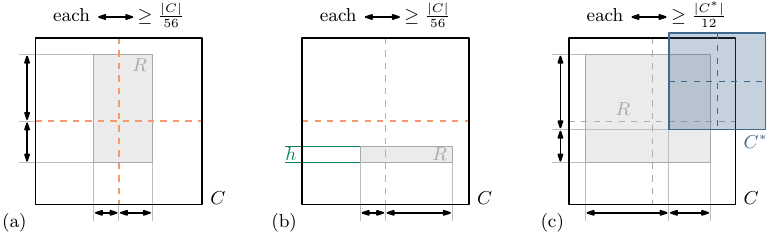}
\caption{The no-thin-slices invariant when (a) both midlines cut the range and (b) only the vertical midline does. In (b), the height $h$ may be arbitrarily small. (c) The compressed guard invariant for a compressed node $(C,R)$ with child $(C^*,R^*)$.}
\label{fig:quadtree-invariants}
\end{figure}

To obtain the packing property, we maintain two invariants for our aligned quadtree.
The first invariant prevents a vertical split from creating a range whose width is too small compared to the cell's side length, and similarly prevents a horizontal split from creating a range whose height is too small.
The second invariant imposes a similar condition for compressed nodes (see Figure~\ref{fig:quadtree-invariants} for an illustration of the invariants).

\begin{invariant}[No thin slices]
\label{inv:slices}
For a regular node $(C,R)$, if the vertical midline of $C$ cuts the interior of $R$, then it divides $R$ into rectangles of width at least $\fixedDelta|C|$ each.
If the horizontal midline of $C$ cuts the interior of $R$, it divides $R$ into rectangles of height at least $\fixedDelta|C|$ each.
\end{invariant}

\begin{invariant}[Compressed guard]
\label{inv:guard}
Let $(C,R)$ be a compressed node with child $(C^*,R^*)$.
Every vertical line through a side of $C^*$ that cuts the interior of $R$ divides $R$ into two rectangles of width at least $\fixedEpsilon|C^*|$ each.
Every horizontal line through a side of $C^*$ that cuts the interior of $R$ divides $R$ into two rectangles of height at least $\fixedEpsilon|C^*|$ each.

We say that $C^*$ is \emph{guarded with respect to $R$}.
\end{invariant}

In the following sections, we prove our main theorem using these invariants.

\begin{theorem}[Main theorem]
\label{thm:main}
We can maintain an aligned quadtree $T_B(P)$ with $O(n)$ nodes and $O(\log n)$ update time under point insertions and deletions, where $n$ is the current number of points in $P$. The aligned quadtree satisfies the quadtree packing property
(Definition~\ref{def:packing}).
\end{theorem}

\subsection{Inserting a point}\label{sec:insertions}

We show how to maintain a linear-size quadtree $T_B(P)$ satisfying Invariants~\ref{inv:slices} and~\ref{inv:guard} for a dynamic point set $P$. In this section, we focus on inserting a point $p$ into $P$. In the next section, we discuss the easier case of deleting a point from $P$.
We use the dynamic edge-oracle tree of Appendix~\ref{app:edge-oracles} with
$O(\log n)$ query and update time.
Given a point $p\in B$, the oracle returns either the leaf whose range contains $p$, or the compressed node $(C,R)$ with child $(C^*,R^*)$ for which $p\in R\setminus C^*$.
If the node returned by the oracle already stores a point or is a compressed node, the following lemma shows how to construct a square $W$ that we can use as the cell of a new node in $T_B(P)$. A regular split of this cell separates $p$ from the other point contained in the leaf (or from the entire cell $C^*$ in the compressed case).



\begin{lemma}[Separating square
]
\label{lem:join}
Let $T_B(P)$ be an aligned quadtree satisfying Invariant~\ref{inv:guard}, let $(C,R)$ be a node in $T_B(P)$, and let $p$ be a point in $R$.
If $(C,R)$ is either a leaf storing a point $q$, or a compressed node with child $(C^*,R^*)$ where $p\notin R^*$, then we can construct a square $W$ in $O(1)$ time with the following properties:
\begin{enumerate}[label=(\roman*),noitemsep]
 \item If $(C, R)$ is a leaf, then $W$ contains $p$ and $q$ in different quadrants. 
    Otherwise, one quadrant of $W$ contains $p$ and another contains the entire cell $C^*$.
 \item $W$ is guarded with respect to $R$ (see Invariant~\ref{inv:guard}).
 \item The regular node $(W,R\cap W)$ satisfies Invariant~\ref{inv:slices}.
 \item If $(C, R)$ is a compressed node, let $Q^*$ be the quadrant of $W$ containing $C^*$ and let $R_Q=R\cap Q^*$.
 Then $C^*$ is guarded with respect to $R_Q$.
\end{enumerate}
\end{lemma}

\begin{proof}
We denote by $A$ the object that gets separated from $p$, which is $\{ q \}$ if $(C, R)$ is a leaf and $C^*$ otherwise. 
We denote by $a$ the size of this object, so that $a =0$ or $a = |C^*|$, respectively. 
Let $b$ be the side length of a smallest closed square containing $A$ and $p$.
As $p \notin A$, we may assume by symmetry that $p$ is either to the right or to the left of $A$ and that the smallest vertical strip containing $A$ and $p$ has width $b$. Recall that rectangles are half-open. Thus $p$ must lie strictly to the left of $A$ in the first case, but may lie on the right boundary of $A$ in the second.

We first fix the vertical midline of $W$, then its horizontal midline, and finally its size. 

\subparagraph{The vertical midline.} To determine the vertical midline of $W$, we make a case distinction depending on the gap between $A$ and $p$, which has width $b - a$, see Figure~\ref{fig:separating-square} for the two cases.

\vspace{0.3cm}
\emph{Case 1: $b>(1+\fixedEpsilon[2])a$.}
In this case, we place the vertical midline of $W$ halfway between~$A$ and $p$.
Its distance to both $A$ and $p$ is $
 \frac{b-a}{2}>\frac{1}{2}(b - 1/(1+\frac{2}{12})b)
 =\frac{b}{14}$. 
Since $R$ contains $p$ and a point in~$A$, the midline
divides $R$ into two rectangles of width at least $b/14$.

\emph{Case 2: $b\leq(1+\fixedEpsilon[2])a$.}
In this case, $a >0$ and thus $A$ must be the square $C^*$.
We place the vertical midline of $W$ on the right or left boundary of $A$, depending on if $p$ is to the right or strictly left of $a$.
This separates $p$ from $A$ even when $p$ lies on the line,
since $A$ does not contain its right boundary.
Since $R$ contains $p$ and a point of $A$, the line intersects
the interior of $R$.
By the \hyperref[inv:guard]{compressed guard invariant}, the line through the chosen side of $A = C^*$ divides $R$ into
two rectangles of width at least $\fixedEpsilon a\geq
\fixedEpsilon b/(1+2\fixedEpsilon)=b/14$.

\emph{In both cases:}
The chosen vertical midline lies between $p$ and $A$ and divides $R$ into two rectangles of width at least $b/14$. Furthermore, every point of $A$ and the point $p$ lie within horizontal distance $b$ of the midline.

\begin{figure}[b]
\centering
\includegraphics[]{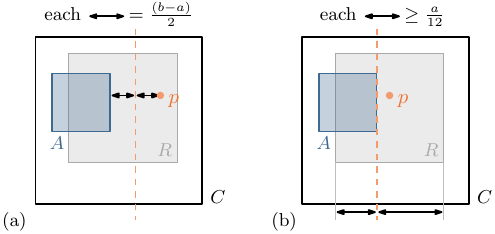}
\caption{The placement of the vertical midline (orange dashed) in (a) Case 1 and (b) Case~2.}
\label{fig:separating-square}
\end{figure}

\subparagraph*{The horizontal midline.}
We next choose the horizontal midline of $W$.
We place it at or below the bottom of $A$.
If $A=\{q\}$, its bottom simply means the height of $q$.
The choice depends on how far the bottom of $A$ lies above the bottom of $R$, see Figure~\ref{fig:other-midline}.

\vspace{0.3cm}
\emph{Case 1: the bottom of $A$ lies below or at most $b/6$ above the bottom of $R$.}
We place the horizontal midline at the lower of the bottom of $A$ and $R$. If this places the midline on the bottom of $R$, then the vertical distance from the midline to the bottom of $A$ is at most $b/6$.

\emph{Case 2: the bottom of $A$ lies more than $b/6$ above the bottom of $R$.}
We place the horizontal midline at distance $b/12$ below the bottom of $A$.
By construction, this line divides $R$ into two rectangles of height greater than $b/12$.

\begin{figure}
\centering
\includegraphics[]{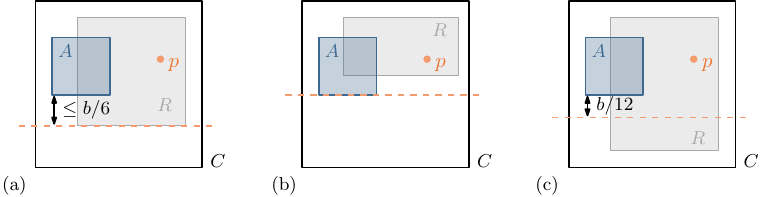}
\caption{The horizontal midline (dashed orange). In Case 1, it passes through the bottom of (a)~$R$ or (b) $A$. (b) In Case 2, the midline lies $b/12$ below $A$.}
\label{fig:other-midline}
\end{figure}

\emph{In both cases:}
The horizontal midline lies at most $b/6$ below the bottom of $A$.
Since $A$ and $p$ fit in a square of side length $b$, their vertical distance to the bottom of $A$ is at most~$b$.
Their vertical distance to the chosen midline is therefore at most $b+b/6=7b/6$.
If this midline cuts $R$, it leaves height greater than $b/12$ on both sides.

\subparagraph{The size of $\mathbf{W}$.} Let $z$ be the intersection point of the horizontal and vertical midline, i.e., the centre point of $W$. We set $W_0$ to be the square centred at $z$ of side length $\frac{5}{2}b$. Next, we apply the following recursive procedure on $W := W_0$ (see Figure~\ref{fig:full-construction}).
If $W$ is guarded with respect to $R$, we return $W$.
Otherwise, we choose a line $\ell$ through a side of $W$ that violates the guarding property with respect to $R$. If $\ell$ goes through the left side of $W$, then let $d$ be the horizontal distance from $\ell$ to the left side of $R$. We analogously define $d$ when it passes through another side of $W$. We move all sides of $W$ outwards by $d$ by setting the width~$w$ of~$W$ to be $w+2d$, and recurse. Note that after this enlargement, this side of~$W$ will never trigger another enlargement, as the line through this side no longer properly intersects $R$. Thus, we obtain a square $W$ that is guarded with respect to~$R$ after at most four enlargements.

Next, we bound the size of the resulting $W$. By construction, the horizontal midline intersects $R$, while the vertical midline either intersects $R$, or lies at the bottom of $A$.
In the last case the vertical midline lies less than
$a\leq b$ below $R$ because $A$ intersects $R$.
A line through a vertical (resp. horizontal) side of $W$ that cuts $R$ therefore leaves
width (resp. height) at least $w/2-b$ towards~$z$.
Since $w\geq \frac{5}{2}b$, $
 w/2-b\geq \fixedEpsilon w$.
The guarding property can therefore fail only on the outward side. So, if the guarding property fails we have that $d < \fixedEpsilon w$.
Each enlargement increases
the side length to $
 w+2d<(1+2\fixedEpsilon)w=\frac76w$.
 Since there are at most four recursive calls that increase the side length by a factor $\frac{7}{6}$, the returned square $W$ satisfies $
 \frac{5b}{2}\leq |W|
 \leq \frac{5b}{2}\left(\frac76\right)^4<5b$.

\begin{figure}[]
\centering
\includegraphics[]{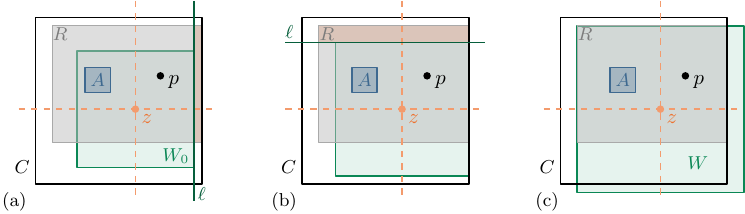}
\caption{Two steps of enlargement are needed from $W_0$ to reach the final square $W$. In each step the line $\ell$ and the corresponding orange rectangle violate that $W$ is guarded with respect to $R$.}
\label{fig:full-construction}
\end{figure}

\subparagraph*{Property (i).}
Every point of $A$ and the point $p$ lie within horizontal
distance $b$ and vertical distance $7b/6$ of vertical and horizontal midline, respectively.
Thus $W$ contains both~$A$ and~$p$. The vertical midline separates~$A$ and~$p$, while they lie on the same side of the vertical midline. Thus, $A$ and $p$ are contained in different quadrants of $W$.

\subparagraph*{Property (ii).} By construction, the square $W$ is guarded with respect to $R$.

\subparagraph*{Property (iii).}
We must show that the vertical midline of $W$, if it cuts
the interior of $W\cap R$, leaves two rectangles of width
at least $\fixedDelta w$.
The horizontal midline must similarly leave two rectangles
of height at least $\fixedDelta w$.
We argue first for the vertical midline.
We distinguish whether the initial square $W_0$ (of side
length $5b/2$) lies entirely in $R$.

If $W_0\subseteq R$, then $R\cap W$ still contains $W_0$,
since we only enlarged the square.
Both midlines pass through the centre of $W_0$, so each
leaves width or height at least $5b/4$ on both sides in $R\cap W$.
Since $w<5b$, this is greater than the required $\fixedDelta w$.

If $W_0\not\subseteq R$, at least one side of $W_0$ already
lies outside $R$ and moves farther away as we enlarge.
Only the other three sides can then trigger an enlargement, so $
 w\leq\frac{5b}{2}\left(\frac76\right)^3<4b$.
By construction, the vertical midline divides $R$ into two
rectangles of width at least $b/14$. As the width of $W_0$ is $5b/2$ the intersection $R \cap W$ still leaves rectangles of width at least $b/14$. Using that $w < 4b$, we find the width is at least $b/14 > \fixedDelta w$.

For the horizontal midline, if it cuts the interior of $R\cap W$, the same argument applies with heights instead of widths: its construction leaves height greater than $b/12\geq b/14$ on both sides in $R$.
Thus both midlines satisfy Invariant~\ref{inv:slices}, proving property~(iii).

\subparagraph*{Property (iv).} Property (iv) states that if $(C, R)$ is a compressed node, then $C^*$ is guarded with respect to $R_Q$, where  $R_Q=R\cap Q^*$ for $Q^*$ the quadrant of $W$ containing $C^*$. See Figure~\ref{fig:property_iv}.
Suppose that $(C,R)$ is indeed a compressed node, and so $A=C^*$. Recall that~$a=|C^*|$.


Let $\ell$ be a vertical line through a side of $C^*$ that
cuts the interior of $R_Q$ (see Figure~\ref{fig:property_iv}).
We show that $\ell$ divides $R_Q$ into two rectangles
of width at least $\fixedEpsilon a$.
Each vertical side of $R_Q$ coincides with (part of) a side of $R$,
a side of $W$, or the vertical midline of $W$.
If a side of $R_Q$ lies on a side of $R$, its distance to $\ell$ is
at least $\fixedEpsilon a$, as $C^*$ is guarded with respect to $R$ by Invariant~\ref{inv:guard}.
If a side of $R_Q$ lies on a side of $W$, its distance to $\ell$ is
at least $5b/4-b= b/4\geq\fixedEpsilon a$ by construction of $W$.
It remains to consider the case where a side of~$R_Q$ coincides with the vertical midline of $W$.
In Case 1 of the construction of this vertical midline, we placed it at distance greater than $\fixedEpsilon a$ from $C^*$.
In Case 2, we placed it on a side of $C^*$.
The line through this side lies on the boundary of $R_Q$,
so it does not cut its interior.
The line through the opposite side of $C^*$ is at distance
$a\geq\fixedEpsilon a$ from this boundary.
Thus, both rectangles have the required width.

\begin{figure}
    \centering
    \includegraphics{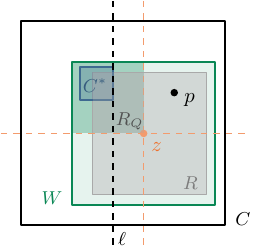}
    \caption{The square $C^*$ must be guarded with respect to $R_Q$. The quadrant $Q^*$ is highlighted.}
    \label{fig:property_iv}
\end{figure}

For horizontal lines, only a side of $R_Q$ that lies on the
horizontal midline of $W$ needs a separate argument. (If a side of $R_Q$ lies on a side of $W$, its distance to a horizontal line through a side of $C^*$ is
at least $5b/4-7b/6= b/12\geq\fixedEpsilon a$ by construction of $W$.)
By construction, this midline lies at the bottom of $R$,
at the bottom of $C^*$, or $b/12$ below $C^*$.
If a side of $R_Q$ lies at the bottom of $R$, the required distance follows
because $C^*$ is guarded with respect to $R$ by
Invariant~\ref{inv:guard}.
If it lies at the bottom of $C^*$, the line through that side
does not cut the interior of $R_Q$, and the line through the
top side is at distance $a\geq\fixedEpsilon a$ from the midline.
Otherwise, both horizontal sides of $C^*$ lie at distance
at least $b/12\geq\fixedEpsilon a$ above the midline.
Thus every horizontal line through a side of $C^*$ that cuts
$R_Q$ leaves two rectangles of height at least $\fixedEpsilon a$.

\subparagraph{Complexity.}
Each choice uses a constant number of arithmetic operations and comparisons, and there are at most four enlargements.
The construction therefore takes $O(1)$ time.
\end{proof}
To insert a point $p$ into $P$, we update the tree $T_B(P)$ as follows. Recall that given a point $p\in B$, the oracle returns either the leaf whose range contains $p$, or the compressed node $(C,R)$ with child $(C^*,R^*)$ for which $p\in R\setminus C^*$. Based on this we make a case distinction.

\subparagraph*{Case 1: the oracle returns a leaf $(C,R)$.}
If the leaf is empty, we store $p$ there.
Otherwise, let $q$ be its stored point.
We apply Lemma~\ref{lem:join} to obtain a square $W$ whose quadrants separate $p$ from $q$, and let $V=R\cap W$.
We create a regular node $y=(W,V)$.
For each quadrant $Q$ of $W$, we give $y$ a leaf child $(Q,V\cap Q)$.
We store $p$ and $q$ in their respective leaves; the other two leaves remain empty.
Finally, we make $(C, R)$ a compressed node with $y$ as its only child.

\begin{figure}
\centering
\includegraphics[]{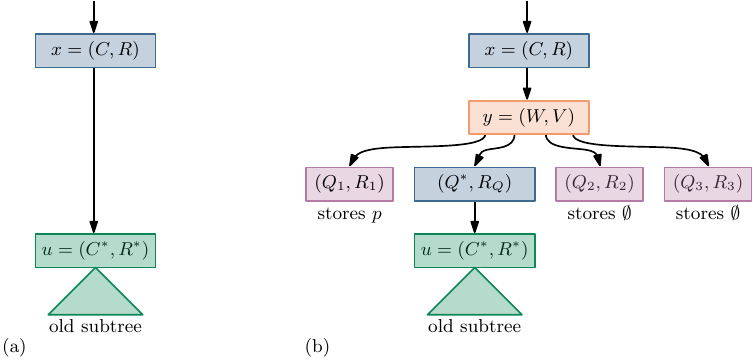}
\caption{Case 2 of the insertion algorithm: the subtree (a) before and (b) after inserting $p$. Compressed nodes are blue, regular nodes are orange, and leaves are purple. The green node $u$ (and its subtree) can be any type of node.}
\label{fig:insertion}
\end{figure}

\subparagraph*{Case 2: the oracle returns a compressed node $(C,R)$.} Figure~\ref{fig:insertion} illustrates the construction.
Let $u=(C^*,R^*)$ be its child.
The node is returned because $p\in R\setminus C^*$.
We apply Lemma~\ref{lem:join} to obtain a square $W$ with $p$ in one quadrant and $C^*$ in another.
We create a regular node $y=(W,V)$ where $V=R\cap W$.
Let $Q^*$ be the quadrant of $W$ containing $C^*$, and let $R_Q=V\cap Q^*=R\cap Q^*$.
We give $y$ a compressed child $(Q^*,R_Q)$, whose only child is the old node $u$.
By Lemma~\ref{lem:join}(i), $R_Q\cap C^*=R^*$, so this preserves the range of $u$.
For each of the other three quadrants $Q_i$ we give $y$ a leaf child $(Q_i,R_i = V\cap Q_i)$.
We store $p$ in its quadrant leaf and leave the other two leaves empty.
Finally, we replace the child of $(C,R)$ by $y$.

\begin{lemma}[Insertion correctness]
\label{lem:insertion}
The insertion algorithm produces an aligned quadtree storing $P\cup\{p\}$, and preserves Invariants~\ref{inv:slices} and~\ref{inv:guard}.
Given the node returned by the edge-oracle tree when searching for $p$, it changes $O(1)$ nodes in $T_B(P)$ and takes $O(1)$ time.
\end{lemma}

\begin{proof}
Inserting into an empty leaf preserves all properties immediately.
Otherwise, we first show that the updated tree is a valid aligned quadtree storing $P\cup\{p\}$.
In both cases of the insertion algorithm above, the new regular node $y=(W,V)$ has the required quadrant children, and each new compressed link has a child with non-empty range equal to its cell intersected with the parent range.
For the link to the old child in Case 2 of the insertion algorithm,  Lemma~\ref{lem:join}(i) implies that  $R_Q\cap C^*=R^*$.
In Case 1 of the insertion algorithm, $p$ and $q$ are stored in different quadrant leaves.
In Case 2, the old subtree keeps its points and ranges, and $p$ is stored in a different quadrant.
All other nodes stay unchanged, so every point is stored exactly once in a leaf whose range contains it.

\subparagraph*{Invariants and running time.}
Lemma~\ref{lem:join} property~(ii) states that $W$ is guarded with respect to $R$, and thus the compressed guard invariant still holds for $(C,R)$, and property~(iii) gives the no-thin-slices condition at $y = (W,V)$.
In Case 2 of the insertion algorithm, property~(iv) states that $C^*$ is guarded  with respect to $R_Q$, and thus the compressed guard invariant also holds for $(Q^*,R_Q)$.
All other regular and compressed nodes keep the cells, ranges, and children used by their invariants.
The construction of $W$ takes $O(1)$ time by Lemma~\ref{lem:join}.
Both insertion cases then create or change $O(1)$ nodes and child links, and compute $O(1)$ rectangle intersections.
Thus insertion takes $O(1)$ time once node found for $p$ is known.
\end{proof}

\subsection{Deleting a point}
\label{sec:deletions}

To support deletions, we simply use the global rebuilding
technique of Overmars~\cite[Chapter~5]{Overmars1983}.
To delete a point, we locate its leaf and remove the point, leaving
all nodes, cells, and ranges unchanged.
This preserves both invariants, but may leave empty subtrees.
To maintain linear size, we rebuild in \emph{rebuild phases}.
We maintain $P$ in a balanced search tree, ordered
by the time each point was inserted.
If a phase starts with fewer than eight points, we simply rebuild the aligned quadtree.
Otherwise, we set $N=|P|$ at the start of the phase.
The phase lasts $\lceil N/4\rceil$ updates, and so throughout the phase the size of $P$, which is $n$, remains in $\Theta(N)$. 
We answer queries using the current tree $T$,
which we keep updating, whilst building a fresh aligned quadtree $T'$ as follows.

At the start of the phase, we record the time $t$ at which a point was last inserted in $P$.
Over the course of this phase, we scan the balanced search tree in insertion order from the start to time $t$ and insert these point into $T'$.
After each update, we insert the next four surviving points into $T'$, or all remaining points if fewer than four remain up to time $t$.
We record the insertion time of the last point inserted into $T'$, so that after the next update we can continue the scan, even if that point is deleted.

New insertions during the phase are performed in $T$, $T'$, and the balanced
search tree.
When a point is deleted during the phase, we remove the point from $T$, from $T'$ if present, and from the balanced search tree.
Thus $T'$ always stores exactly the surviving points already
passed by the scan or inserted during the phase.
There are at most $N$ points in the balanced search tree up to time $t$, and we insert 4 of these points at each update, so we finish processing these point by the end of the phase.
After $\lceil N/4\rceil$ updates, $T'$ therefore stores exactly
$P$. We make $T'$ the active tree and begin the next rebuild phase. 
Each fresh tree receives $O(N)$ insertions and remains active
for only the following phase, so all maintained trees have
$O(N)=O(n)$ nodes.
Finally, we note that after replacing $T$ by $T'$, one still has to  `clean up' memory. Since $T$ has $O(N')$ size where $N' = |P|$ at the end of the rebuild phase,  we simply do constant work to remove $T$ from memory for the next  $\lceil N'/4\rceil$ updates. 
Each update performs constantly many quadtree updates and balanced-search-tree operations. By Lemma~\ref{lem:insertion} and the edge oracle of Appendix~\ref{app:edge-oracles}, these take $O(\log N)=O(\log n)$ time.

\begin{theorem}[Dynamic maintenance]
\label{thm:dynamic}
For any fixed bounding box $B$, we can maintain an aligned quadtree $T_B(P)$ that satisfies Invariants~\ref{inv:slices} and~\ref{inv:guard} with $O(n)$ nodes and $O(\log n)$ update time under point insertions and deletions, where $n$ is the current number of points in $P$.
\end{theorem}

\subsection{Quadtree packing}\label{sec:quadtree_packing}

It remains to show that Invariant~\ref{inv:slices} implies the quadtree packing property of Definition~\ref{def:packing}.

\begin{lemma}[Quadtree packing]
\label{lem:antichain}
An aligned quadtree satisfying the no-thin-slices condition of Invariant~\ref{inv:slices} has the quadtree packing property.
\end{lemma}

\begin{proof}
Let $D=D(x,r)$ be a disk with centre $x$ and radius $r>0$, and let $\rho>0$.
Consider a set $\Pi$ of nodes as in Definition~\ref{def:packing}: their ranges intersect or touch $D$, no node is an ancestor of another, and the lowest common ancestor of any two has a cell of side length at least~$\rho$.
We must upper bound the number of nodes in $\Pi$.
To this end, we construct for each node $(C,R)\in\Pi$ a square $S_C$ of side length $m=\fixedDelta\rho$, where all squares $S_C$ are interior-disjoint. 
The square $S_C$ need not lie in $R$.
We then bound $|\Pi|$ using the area of a slightly larger disk.

\subparagraph*{Constructing the squares.}
Fix a node $(C,R)\in\Pi$ and choose a point $p_C$ where $R$ intersects or touches $D$. We construct $S_C$ around this point as follows. 
We traverse the path from the root to $(C,R)$.
Among the regular nodes before $(C,R)$, we consider those whose cell has side length at least $\rho$ and whose vertical midline cuts the interior of their range.
Suppose there is such a node, and let $(A,U)$ be the last one on this path.
Let $Q$ be the quadrant of $A$ that we follow towards $(C,R)$.
By Invariant~\ref{inv:slices}, the child range $Q\cap U$ has width at least $\fixedDelta|A|\geq m$.
Moreover, it contains $R$, since a child's range is always contained in its parent's range.

\begin{figure}
    \centering
    \includegraphics{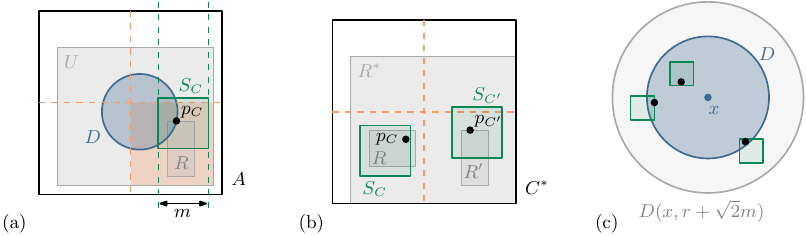}
    \caption{Constructing and counting the squares $S_C$. (a) The range $Q \cap U$ (orange) has width at least $m$, the vertical lines are using to choose the vertical sides of $S_C$. (b) The lowest common ancestor $(C^*,R^*)$ of $(C,R)$ and $(C,R')$ separates $R$ and $R'$ vertically. (c) The squares $S_C$ and the corresponding points $p_C \in D$. Each $S_C$ is contained in $D(x,r+\sqrt{2}m)$.}
    \label{fig:packing_property proof}
\end{figure}

We can therefore place two vertical lines at distance $m$ apart, between the left and right sides of $Q\cap U$, with $p_C$ between or on these lines.
These lines determine the left and right sides of $S_C$, see Figure~\ref{fig:packing_property proof}(a).
If there is no such node $(A,U)$, we simply choose any two vertical lines at distance $m$ apart with $p_C$ between or on them.
We choose the bottom and top sides in the same way, using horizontal midlines and heights instead of vertical midlines and widths.
The four chosen lines bound a closed square $S_C$ of side length $m$ containing $p_C$.

\subparagraph*{The squares are interior-disjoint.}
Consider two nodes $(C,R)$ and $(C',R')$ in $\Pi$.
Their lowest common ancestor $(C^*,R^*)$ is regular, with
$|C^*|\geq\rho$.
One of its midlines separates~$R$ and $R'$ and thus cuts the interior of $R^*$, see Figure~\ref{fig:packing_property proof}(b). We may assume it is vertical, with $R$ to its left and $R'$ to its right.
When constructing $S_C$, we therefore choose a vertical cut
at $(C^*,R^*)$ or farther down the path to $(C,R)$.
The child range used in this construction stays to the left
of the separating midline, so $S_C$ does too.
Similarly, $S_{C'}$ stays to its right.
The squares may touch the midline, but their interiors do not
overlap.

\subparagraph*{Counting the squares.}
Each square $S_C$ contains a point $p_C$ in $D$ and has side length $m$.
Every point of $S_C$ is therefore at distance at most $\sqrt2m$ from $p_C$.
All squares lie in the disk $D(x,r+\sqrt2m$, see Figure~\ref{fig:packing_property proof}(c).
Since the square are interior-disjoint and each has area~$m^2$, we obtain
$
 |\Pi|m^2\leq\pi(r+\sqrt2m)^2$.
Substituting $m=\fixedDelta\rho$ gives, which proves the packing property with $M =  56^2 \pi$.
\begin{equation*}
 |\Pi|\leq\pi\left(\frac{r}{\fixedDelta\rho}+\sqrt2\right)^2
 \leq 56^2 \pi\left(1+\frac r\rho\right)^2.\qedhere
 \end{equation*}
\end{proof}
By Theorem~\ref{thm:dynamic}, we maintain a linear-size aligned quadtree satisfying both invariants in the stated update time.
Together with Lemma~\ref{lem:antichain}, we obtain our main theorem for a fixed bounding box: Theorem~\ref{thm:main}.

\section{Extensions}\label{app:extension}\label{sec:extension}

We first extend our quadtree to points in $\R^d$ for any fixed $d$.
We then show how to grow the root when a point is inserted outside it, i.e., we lift the assumption that $B$ is a fixed bounding box.
In $\R^d$, cells are axis-aligned hypercubes and ranges are hyperboxes. For readability, we drop the ``hyper'' from here on.
A regular split
divides a cube into $2^d$ equal-sized subcubes. Each box is half-open in the sense that it includes its lower face and
excludes its upper face in every coordinate. We generalise Invariants~\ref{inv:slices} and \ref{inv:guard}: 
For the no-thin-slices invariant, we
replace $\fixedDelta$ by a constant $\delta_d$ determined below.
For the compressed-guard invariant, we still require width $\fixedEpsilon|C^*|$
on either side of a face-supporting plane of the child cube $C^*$ whenever
that plane cuts the interior of the parent range. We recreate the cube $W$ that maintains these invariants, which is the higher dimensional equivalent of Lemma~\ref{lem:join}:

\begin{lemma}[Separating box]\label{lem:d-join}
Let $R$ be a box and let $A$ be either a point in $R$ or a cube that
intersects $R$ and is guarded with respect to $R$. Given $p\in R\setminus A$,
we can construct in $O_d(1)$ time a cube $W$ with $A$ in one subcube $Q$
and $p$ in another of its regular split. The cube $W$ is guarded with respect to $R$, and
$(W,R\cap W)$ satisfies the no-thin-slices invariant. If $A$ is a cube,
it is also guarded with respect to $R\cap Q$.
\end{lemma}

\begin{proof}
Let $a$ be the side length of $A$, with $a=0$ if $A$ is a point.
We choose the midplanes of $W$ as in Lemma~\ref{lem:join}: Draw the smallest closed box containing $A$ and $p$, and let $b$ be
the length of its longest side. We first decide in which coordinate to place the first midplane. If $b>a$, we take the first coordinate
parallel to this side. In this coordinate, $A$ occupies an interval of
length $a$ and $p$ extends it to length $b$. If $b=a$, then $p$ lies
on an excluded face of $A$, and we take the first coordinate perpendicular
to that face. In both cases the gap between $A$ and $p$ in the first
coordinate is $g=b-a$.
If $g>a/6$, we place the first midplane halfway across the gap $g$.
Since $R$ contains $p$ and a point of $A$, the midplane divides $R$
into two boxes of width at least $g/2>b/14$ in the first coordinate.
Otherwise, we place it on the face of $A$ facing $p$. This separates
them even when $p$ lies on the excluded face. Since $R$ contains $p$ and intersects $A$, this plane cuts the interior
of $R$. Because $A$ is guarded with respect to $R$,  this cut gives width at least $a/12\ge b/14$ on either side.

\subparagraph{The remaining midplanes.}
In each other coordinate we use the horizontal midline construction of
Lemma~\ref{lem:join}. In particular, view this coordinate as height: If the bottom of $A$
lies more than $b/6$ above the bottom of $R$, we place the midplane $b/12$
below $A$. Otherwise, we place it at the lower of the two bottoms.
As in the planar proof, any such midplane that cuts $R$ leaves height
at least $b/12$ on both sides. It lies at most $b/6$ below $A$, so every
point of $A$ and the point $p$ lie within distance $7b/6$ of the midplane.

\subparagraph{The size of $\mathbf{W}$.}
We centre $W$ at the common intersection $z$ of all the midplanes. We set $W_0$ to be the cube with centre $z$ and side length
$5b/2$. Since $5b/4>7b/6$, this cube contains $A$ and $p$ in different subcubes. As before, we apply a recursive procedure on $W := W_0$ that enlarges $W$ in each step until $W$ is guarded with respect to $R$.
A midplane outside $R$ passes through the bottom of $A$ in its coordinate.
Since $A$ intersects~$R$, this midplane is at distance at most $a\le b$ from $R$. Hence, if a plane through a face of $W$
cuts $R$, it leaves width at least $|W|/2-b\ge |W|/12$ towards the centre.
If the width on the other side is $h<|W|/12$, and thus $W$ is not guarded with respect to $R$, we move every face outwards
by $h$. The offending face (the one that triggered this enlargement) reaches the boundary of $R$ and cannot trigger
another enlargement. There are $2d$ faces, so the final side length is at
most $(5b/2)(7/6)^{2d}$ and the construction takes $O(1)$ time for fixed $d$.

\subparagraph{$\mathbf{(W,R\cap W)}$ satisfies the no-thin-slices invariant.} Consider a midplane of $W$ that cuts the interior of $R\cap W$.
On either side of the midplane, the boundary of $R$ is at distance
at least $b/14$ in the coordinate perpendicular to the midplane.
The boundary of $W$ is at distance $|W|/2\ge 5b/4$ in the same
direction. The nearer of these two boundaries determines the width
of $R\cap W$ on that side, so this width is at least $b/14$.
It remains to express this width in terms of the final side length
of~$W$. Since $|W|\le (5b/2)(7/6)^{2d}$, we have
$b/14\ge \frac{1}{35}(6/7)^{2d}|W|$.
Thus the no-thin-slices invariant (Invariant~\ref{inv:slices}) holds with $\delta_d=(6/7)^{2d}/35$.

\subparagraph{$\mathbf{A}$ is guarded with respect to $\mathbf{R\cap Q}$.} Fix a coordinate $i$ and let
$H$ be the hyperplane containing a face of $A$ perpendicular to the
$i$th coordinate axis. If $H$ cuts the interior of $R\cap Q$, we must
show that it leaves width at least $a/12$ on each side in coordinate $i$.
If a face of $R\cap Q$ parallel to $H$ lies on a face
of $R$, its distance to $H$ is at least $a/12$, since $A$ is guarded
with respect to $R$.
If the face lies on a face of $W$, its distance to $H$ is at least
$5b/4-7b/6=b/12\geq a/12$. It remains to consider a face that lies
on a midplane of $W$ and cuts the interior of $R$. By construction,
this midplane is either at distance at least $a/12$ from~$A$ or
passes through a face of $A$. In the latter case, $H$ cannot be the
midplane, since $H$ cuts the interior of $R\cap Q$. Thus $H$ passes
through the opposite face of $A$, at distance $a$ from the midplane.
Hence $H$ leaves width at least $a/12$ on both sides in $R\cap Q$ 
as required.
\end{proof}

We generalise the packing packing property to dimension $d$ as follows.

\begin{definition}[Higher dimensional quadtree packing property]
\label{def:packing_d_dim}
For fixed $d$, an aligned quadtree in $\R^d$ has the \emph{quadtree packing property}
if there is a constant $M_d$ with the following guarantee.
Let $D$ be any disk of radius $r>0$, and let $\rho>0$.
Let $\Pi$ be a set of quadtree nodes whose ranges intersect or
touch~$D$ such that (1)  no node in $\Pi$ is an ancestor of another, and  (2) 
the cell of the lowest common ancestor of any two nodes in $\Pi$
has side length at least $\rho$.
Then $\Pi$ contains at most $ M_d\left(1+\frac r\rho\right)^d$ nodes. 
\end{definition}

\begin{lemma}[Packing in fixed dimension]\label{lem:d-packing}
For fixed $d$, an aligned quadtree in $\R^d$ satisfying the no-thin-slices invariant with constant
$\delta_d$ satisfies the quadtree packing property.
\end{lemma}

\begin{proof}
Let $D$ be a ball of radius $r$, and let $\Pi$ be a set of nodes as in
Definition~\ref{def:packing_d_dim}. For each $(C,R)\in\Pi$, choose a point $p_C$
where $R$ intersects or touches $D$. In Lemma~\ref{lem:antichain} we chose
the horizontal and vertical intervals of a square around $p_C$ independently. We make the
same choice in each of the $d$ coordinates to obtain a cube of side length
$\delta_d\rho$.

For example, to choose its first interval:
Follow the path from the root to $(C,R)$, stopping just before $(C,R)$.
At each regular node $(A,U)$ on this path, consider the midplane of $A$
perpendicular to the first coordinate axis. Consider all the nodes for which
$|A|\ge\rho$ and this midplane cuts the interior of $U$. Choose the
one closest to $(C,R)$.
By the no thin slices invariant, $U$
has width at least $\delta_d\rho$ in the first coordinate. Since $R\subseteq U$,
we can place a closed interval of length $\delta_d\rho$ between the two ends of $U$ so that it contains the first coordinate of $p_C$. If there is no such ancestor, any interval of
length $\delta_d\rho$ containing this coordinate suffices.

We show that the cubes constructed for any two distinct nodes $x,y\in\Pi$
have disjoint interiors. Let $u$ be their lowest common ancestor.
The paths from $u$ to $x$ and $y$ leave $u$ through different children.
Hence some midplane of $u$ separates these two child ranges and therefore
also separates the ranges of $x$ and $y$.
By renaming the coordinates, assume that this midplane $H$ is perpendicular
to the first coordinate axis. We may assume that the path to $x$
goes to the left of $H$ and the path to $y$ goes to the right.
It suffices to show that the cube for $x$ stays to the left of $H$
and the cube for $y$ stays to its right.

Consider how we chose the first-coordinate interval for $x$. The cell
of $u$ has side length at least $\rho$ by the assumption on $\Pi$.
Moreover, $H$ cuts the interior of the range of $u$, since this range
contains a non-empty child range on each side of $H$. Thus $u$ is one
of the ancestors we consider when choosing this interval.
We use the last such ancestor on the path to $x$. If we choose $u$, we place the interval between the left and right
boundaries of the child range on the path to $x$.
If we use a later
ancestor, its child range is contained in that same left child range,
since ranges are nested along the path. In either case, the interval
does not extend to the right of $H$. Hence the entire cube for $x$
stays to the left of $H$. The same argument shows that the cube for
$y$ stays to the right of $H$. Their interiors are therefore disjoint.

It remains to count the cubes. Each cube has side length $m$ and
contains a point of $D$. Every point of the cube is within distance
$\sqrt{d}\,m$ of that point, so all cubes lie in the ball with the
same centre as $D$ and radius $r+\sqrt{d}\,m$. Let $v_d$ denote the
volume of the unit ball in $\R^d$. Since the cubes have disjoint
interiors, their total volume satisfies
$|\Pi|m^d\le v_d(r+\sqrt{d}\,m)^d$.
Dividing by $m^d$ and substituting $m=\delta_d\rho$ gives
$|\Pi|\le v_d(r/(\delta_d\rho)+\sqrt{d})^d
\leq M \cdot (1+r/\rho)^d)$ for $M = v_d/\delta_d^d + \sqrt{d}$.
\end{proof}

\begin{theorem}[Fixed dimension]\label{thm:extensions}
For fixed constant $d$ and a fixed bounding cube, an aligned quadtree for $n$ points
in $\R^d$ uses $O_d(n)$ space and supports updates and point location in
$O(\log n)$ time. It satisfies the quadtree packing property of Definition~\ref{def:packing_d_dim}.
\end{theorem}

\begin{proof}
We use the insertion algorithm of Section~\ref{sec:insertions}, giving
each regular node $2^d$ children. Lemma~\ref{lem:d-join} supplies the cube
$W$ that separates the new point from the old point or the old child cell.
Thus an insertion preserves both invariants and changes $O(1)$ nodes and
edges once we have located the new point. The degree is at most $2^d+1$,
so the edge oracle of Appendix~\ref{app:edge-oracles} and the rebuilding
algorithm of Section~\ref{sec:deletions} give the stated time and space
bounds. Lemma~\ref{lem:d-packing} gives the packing property.
\end{proof}

\begin{theorem}[No fixed bounding box]\label{thm:unbounded}
The bounds of Theorem~\ref{thm:extensions} hold without knowing a bounding
cube in advance.
\end{theorem}

\begin{proof}
We start with any (finite) root cube $B$. When we insert a point $p$
outside $B$, we grow the root so that one child contains $B$ and another
contains $p$.
We never need to shrink the root after a deletion, since it still
contains all remaining points and the bounds are independent of its
side length.

\subparagraph{Growing the root.}
Choose a coordinate in which $p$ lies outside $B$, and let $z$ be a
corner of the face of $B$ facing $p$ in that coordinate. Let $b$ be the
side length of a smallest closed cube containing $B$ and $p$.
We centre a new root cube $W$ with range $W$ of side length $4b$ at $z$. Both $B$ and
$p$ lie within distance $b$ of $z$ in every coordinate, so $W$ contains
both. Since $z$ is a corner of $B$, all of $B$ lies in one child cube
$Q$ of $W$. In the chosen coordinate, the midplane through $z$
separates $p$ from $B$, so we store $p$ in a different child leaf.
If the old root was a leaf, we copy its point, if any, to the leaf
$(Q,Q)$. Otherwise, we make $(Q,Q)$ a compressed node with the old
root as its child. Since $Q\cap B=B$, the old root keeps its range,
as does every node below it.

\subparagraph{Preserving the invariants.}
At the new root, each midplane bisects the full range $W$, so
Invariant~\ref{inv:slices} holds. If $(Q,Q)$ is compressed, we must
also show that $B$ is guarded with respect to $Q$. Write $a=|B|$.
In each coordinate, one face of $B$ lies on the boundary of $Q$.
The opposite face divides $Q$ into two boxes of widths $a$ and
$2b-a\ge a$. Both widths are at least $a/12$, as required by
Invariant~\ref{inv:guard}.

If the dimension is constant, changing the root changes $O(1)$ nodes and edges 
For a point location query, if a query point lies outside the current root cube, we report that
it lies outside the tree. During the global rebuilding of Section~\ref{sec:deletions}, we grow
the root of the replacement tree by the same rule whenever we insert
a point outside its root cube. The stated bounds follow.
\end{proof}

\bibliography{references}

@InProceedings{chan2019Locality,
  url = {https://drops.dagstuhl.de/storage/00lipics/lipics-vol124-itcs2019/LIPIcs.ITCS.2019.21/LIPIcs.ITCS.2019.21.pdf},
  author =	{Chan, Timothy M. and Har-Peled, Sariel and Jones, Mitchell},
  title =	{On Locality-Sensitive Orderings and Their Applications},
  booktitle =	{10th Innovations in Theoretical Computer Science Conference (ITCS 2019)},
  pages =	{21:1--21:17},
  series =	{Leibniz International Proceedings in Informatics (LIPIcs)},
  ISBN =	{978-3-95977-095-8},
  ISSN =	{1868-8969},
  year =	{2019},
  volume =	{124},
  editor =	{Blum, Avrim},
  publisher =	{Schloss Dagstuhl -- Leibniz-Zentrum f{\"u}r Informatik},
  address =	{Dagstuhl, Germany},
  doi =		{10.4230/LIPIcs.ITCS.2019.21}
}

@book{HarPeled2011,
  author    = {Sariel Har-Peled},
  title     = {Geometric Approximation Algorithms},
  series    = {Mathematical Surveys and Monographs},
  volume    = {173},
  publisher = {American Mathematical Society},
  year      = {2011},
  doi       = {10.1090/surv/173},
}

@article{BuchinLofflerMorinMulzer2011,
  url = {https://pure.tue.nl/ws/portalfiles/portal/3175221/Metis240245.pdf},
  author  = {Kevin Buchin and Maarten L{\"o}ffler and
             Pat Morin and Wolfgang Mulzer},
  title   = {Preprocessing Imprecise Points for {Delaunay}
             Triangulation: Simplified and Extended},
  journal = {Algorithmica},
  volume  = {61},
  number  = {3},
  pages   = {674--693},
  year    = {2011},
  doi     = {10.1007/s00453-010-9430-0}
}

@book{Overmars1983,
  author = {Mark H. Overmars},
  title = {The Design of Dynamic Data Structures},
  series = {Lecture Notes in Computer Science},
  volume = {156},
  publisher = {Springer},
  year = {1983},
  doi = {10.1007/BFb0014927}
}

@inproceedings{Hoog2024SquareConnectivity,
  url = {https://drops.dagstuhl.de/storage/00lipics/lipics-vol306-mfcs2024/LIPIcs.MFCS.2024.63/LIPIcs.MFCS.2024.63.pdf},
  author = {van der Hoog, Ivor and Nusser, Andr{\'e} and Rotenberg, Eva and Staals, Frank},
  title = {Fully-Adaptive Dynamic Connectivity of Square Intersection Graphs},
  booktitle = {49th International Symposium on Mathematical Foundations of Computer Science (MFCS 2024)},
  series = {Leibniz International Proceedings in Informatics (LIPIcs)},
  volume = {306},
  pages = {63:1--63:17},
  publisher = {Schloss Dagstuhl -- Leibniz-Zentrum f{\"u}r Informatik},
  year = {2024},
  doi = {10.4230/LIPIcs.MFCS.2024.63}
}

@misc{deBerg2026DiskSpannerFull,
  author = {de Berg, Sarita and van der Hoog, Ivor and Rotenberg, Eva and Vistisen, Johanne M{\"u}ller and Wong, Sampson},
  title = {A Dynamic $(1+\varepsilon)$-Spanner for Disk Intersection Graphs},
  year = {2026},
  note = {Full version, arXiv:2604.25397v2},
  url = {https://arxiv.org/abs/2604.25397v2}
}

@article{Alstrup2005TopTrees,
  author = {Stephen Alstrup and Jacob Holm and Kristian de Lichtenberg and Mikkel Thorup},
  title = {Maintaining Information in Fully Dynamic Trees with Top Trees},
  journal = {ACM Transactions on Algorithms},
  volume = {1},
  number = {2},
  pages = {243--264},
  year = {2005},
  doi = {10.1145/1103963.1103966},
  url = {https://arxiv.org/abs/cs/0310065}
}

@inproceedings{Chan2002RAM,
  author = {Timothy M. Chan},
  title = {Closest-Point Problems Simplified on the RAM},
  year = {2002},
  url = {https://tmc.web.engr.illinois.edu/ram_soda.pdf},
  booktitle = {Proceedings of SODA}, pages = {472--473}
}

@article{Baumann2024Disks,
  author = {Alexander Baumann and Haim Kaplan and Katharina Klost and Kristin Knorr and Wolfgang Mulzer and Liam Roditty and Paul Seiferth},
  title = {Dynamic Connectivity in Disk Graphs},
  year = {2024},
  url = {https://page.mi.fu-berlin.de/mulzer/pubs/ddiskDCG.pdf},
  journal = {Discrete \& Computational Geometry}, volume = {71}, pages = {214--277}, doi = {10.1007/s00454-023-00621-x}
}

@inproceedings{Henzinger2020MIS,
  author = {Monika Henzinger and Stefan Neumann and Andreas Wiese},
  title = {Dynamic Approximate Maximum Independent Set of Intervals, Hypercubes and Hyperrectangles},
  year = {2020},
  url = {https://drops.dagstuhl.de/storage/00lipics/lipics-vol164-socg2020/LIPIcs.SoCG.2020.51/LIPIcs.SoCG.2020.51.pdf},
  booktitle = {SoCG}, pages = {51:1--51:14}, doi = {10.4230/LIPIcs.SoCG.2020.51}
}

@misc{LofflerSimonsStrashFull,
  author = {Maarten L{\"o}ffler and Joseph A. Simons and Darren Strash},
  title = {Dynamic Planar Point Location with Sub-Logarithmic Local Updates},
  year = {2013},
  url = {https://arxiv.org/abs/1204.4714},
  note = {Full version of the WADS 2013 paper, arXiv:1204.4714}
}

@inproceedings{Hoog2018Smooth,
  author = {Ivor van der Hoog and Elena Khramtcova and Maarten L{\"o}ffler},
  title = {Dynamic Smooth Compressed Quadtrees},
  year = {2018},
  url = {https://drops.dagstuhl.de/storage/00lipics/lipics-vol099-socg2018/LIPIcs.SoCG.2018.45/LIPIcs.SoCG.2018.45.pdf},
  booktitle = {SoCG}, pages = {45:1--45:15}, doi = {10.4230/LIPIcs.SoCG.2018.45}
}

@article{Arya1998ANN,
  author = {Sunil Arya and David M. Mount and Nathan S. Netanyahu and Ruth Silverman and Angela Y. Wu},
  title = {An Optimal Algorithm for Approximate Nearest Neighbor Searching in Fixed Dimensions},
  year = {1998},
  url = {https://www.cse.ust.hk/~arya/pub/JACM.pdf},
  journal = {Journal of the ACM}, volume = {45}, number = {6}, pages = {891--923}, doi = {10.1145/293347.293348}
}

@article{AryaMount2000Range,
  author = {Sunil Arya and David M. Mount},
  title = {Approximate Range Searching},
  year = {2000},
  url = {https://www.cse.ust.hk/faculty/arya/pub/range.pdf},
  journal = {Computational Geometry}, volume = {17}, number = {3--4}, pages = {135--152}, doi = {10.1016/S0925-7721(00)00022-5}
}

@misc{AcarHudson2008,
  author = {Umut A. Acar and Beno{\^i}t Hudson},
  title = {Dynamic Mesh Refinement with Quad Trees and Off-Centers},
  year = {2008},
  url = {https://www.cs.cmu.edu/~bhudson/papers/acar08dynamic.pdf},
  note = {Author manuscript}
}

@article{Acar2013WellSpaced,
  author = {Umut A. Acar and Andrew Cotter and Beno{\^i}t Hudson and Duru T{\"u}rko{\u g}lu},
  title = {Dynamic Well-Spaced Point Sets},
  year = {2013},
  url = {https://condor.depaul.edu/dturkogl/research/well-spaced-cgta.pdf},
  journal = {Computational Geometry}, volume = {46}, number = {6}, pages = {756--773}, doi = {10.1016/j.comgeo.2012.11.007}
}

@article{BennettYap2017,
  author = {Huck Bennett and Chee Yap},
  title = {Amortized Analysis of Smooth Quadtrees in All Dimensions},
  year = {2017},
  url = {https://doi.org/10.1016/j.comgeo.2017.02.001},
  journal = {Computational Geometry}, volume = {63}, pages = {20--39}, doi = {10.1016/j.comgeo.2017.02.001}
}

@phdthesis{Park2013Thesis,
  author = {Eunhui Park},
  title = {Dynamic Data Structures for Geometric Search and Retrieval},
  year = {2013},
  url = {https://api.drum.lib.umd.edu/server/api/core/bitstreams/e3eaba71-759d-422e-854e-5b425c0bf104/content},
  school = {University of Maryland}
}

@misc{Bhore2020MIS,
  author = {Sujoy Bhore and Jean Cardinal and John Iacono and Grigorios Koumoutsos},
  title = {Dynamic Geometric Independent Set},
  year = {2020},
  url = {https://arxiv.org/abs/2007.08643},
  note = {Full version, arXiv:2007.08643}
}

@misc{BhoreChan2025,
  author = {Sujoy Bhore and Timothy M. Chan},
  title = {Fast Static and Dynamic Approximation Algorithms for Geometric Optimization Problems: Piercing, Independent Set, Vertex Cover, and Matching},
  year = {2025},
  url = {https://arxiv.org/abs/2407.20659},
  note = {Full version of the SODA 2025 paper, arXiv:2407.20659}
}

@inproceedings{ChanYu2025,
  author = {Timothy M. Chan and Yuancheng Yu},
  title = {Dynamic Streaming Algorithms for Geometric Independent Set},
  year = {2025},
  url = {https://drops.dagstuhl.de/storage/00lipics/lipics-vol349-wads2025/LIPIcs.WADS.2025.17/LIPIcs.WADS.2025.17.pdf},
  booktitle = {WADS}, pages = {17:1--17:12}, doi = {10.4230/LIPIcs.WADS.2025.17}
}

@inproceedings{ChanSkrepetos2015,
  author = {Timothy M. Chan and Dimitrios Skrepetos},
  title = {Dynamic Data Structures for Approximate Hausdorff Distance in the Word-RAM},
  year = {2015},
  url = {https://tmc.web.engr.illinois.edu/dyn_haus_cccg.pdf},
  booktitle = {CCCG}, note = {Conference version; journal version in Computational Geometry 60 (2017), 37--44}
}

@inproceedings{VenkatMount2014,
  author = {Prayaag Venkat and David M. Mount},
  title = {A Succinct, Dynamic Data Structure for Proximity Queries on Point Sets},
  year = {2014},
  url = {https://www.cccg.ca/proceedings/2014/papers/paper32.pdf},
  booktitle = {CCCG}
}

@inproceedings{FischerHarPeled2005,
  author = {John Fischer and Sariel Har-Peled},
  title = {Dynamic Well-Separated Pair Decomposition Made Easy},
  year = {2005},
  url = {https://www.cccg.ca/proceedings/2005/32.pdf},
  booktitle = {CCCG}, pages = {235--238}
}

@inproceedings{BhoreToth2021Online,
  author = {Sujoy Bhore and Csaba D. T{\"o}th},
  title = {Online Euclidean Spanners},
  year = {2021},
  url = {https://drops.dagstuhl.de/storage/00lipics/lipics-vol204-esa2021/LIPIcs.ESA.2021.16/LIPIcs.ESA.2021.16.pdf},
  booktitle = {ESA}, pages = {16:1--16:19}, doi = {10.4230/LIPIcs.ESA.2021.16}
}

@inproceedings{Bhore2022Online,
  author = {Sujoy Bhore and Arnold Filtser and Hadi Khodabandeh and Csaba D. T{\"o}th},
  title = {Online Spanners in Metric Spaces},
  year = {2022},
  url = {https://drops.dagstuhl.de/storage/00lipics/lipics-vol244-esa2022/LIPIcs.ESA.2022.18/LIPIcs.ESA.2022.18.pdf},
  booktitle = {ESA}, pages = {18:1--18:20}, doi = {10.4230/LIPIcs.ESA.2022.18}
}

@inproceedings{Agarwal2021Similarity,
  author = {Pankaj K. Agarwal and Xiao Hu and Stavros Sintos and Jun Yang},
  title = {Dynamic Enumeration of Similarity Joins},
  year = {2021},
  url = {https://drops.dagstuhl.de/storage/00lipics/lipics-vol198-icalp2021/LIPIcs.ICALP.2021.11/LIPIcs.ICALP.2021.11.pdf},
  booktitle = {ICALP}, pages = {11:1--11:19}, doi = {10.4230/LIPIcs.ICALP.2021.11}
}

@inproceedings{KisfaludiBak2026Steiner,
  author = {S{\'a}ndor Kisfaludi-Bak and Geert van Wordragen},
  title = {Near-Optimal Dynamic Steiner Spanners for Constant-Curvature Spaces},
  year = {2026},
  url = {https://drops.dagstuhl.de/storage/00lipics/lipics-vol367-socg2026/LIPIcs.SoCG.2026.65/LIPIcs.SoCG.2026.65.pdf},
  booktitle = {SoCG}, pages = {65:1--65:17}, doi = {10.4230/LIPIcs.SoCG.2026.65}
}

@inproceedings{Schonhage1979RAM,
 author={Arnold Sch{\"o}nhage},
 title={On the Power of Random Access Machines},
 booktitle={Automata, Languages and Programming (ICALP 1979)},
 editor={Hermann A. Maurer},
 series={Lecture Notes in Computer Science},
 volume={71},
 pages={520--529},
 publisher={Springer},
 year={1979},
 doi={10.1007/3-540-09510-1_42}
}

@misc{GudmundssonHuangWong2023,
 author={Joachim Gudmundsson and Zijin Huang and Sampson Wong},
 title={Approximating the $\lambda$-low-density value},
 year={2022},
 note={Full version of the COCOON 2023 paper; arXiv:2211.07039v1},
 url={https://arxiv.org/pdf/2211.07039v1}
}

@phdthesis{Hoog2022Thesis,
  author = {Ivor Djinn van der Hoog},
  title = {On Modelling, Structuring and Capturing Geometric Information},
  school = {Utrecht University},
  year = {2022},
  doi = {10.33540/1006},
  url = {https://dspace.library.uu.nl/handle/1874/416117}
}

@misc{GaoHarPeled2025LSO,
  author = {Zhimeng Gao and Sariel Har-Peled},
  title = {Near-Optimal Euclidean Locality-Sensitive Orderings},
  year = {2025},
  eprint = {2310.12792},
  archivePrefix = {arXiv},
  note = {Version 4},
  url = {https://arxiv.org/pdf/2310.12792v4}
}

@article{Kaplan2020Voronoi,
  author = {Haim Kaplan and Wolfgang Mulzer and Liam Roditty and Paul Seiferth and Micha Sharir},
  title = {Dynamic Planar {Voronoi} Diagrams for General Distance Functions and Their Algorithmic Applications},
  journal = {Discrete \& Computational Geometry},
  volume = {64},
  number = {3},
  pages = {838--904},
  year = {2020},
  doi = {10.1007/s00454-020-00243-7},
  url = {https://arxiv.org/pdf/1604.03654v4}
}

@phdthesis{Callahan1995Thesis,
  author = {Paul B. Callahan},
  title = {Dealing with Higher Dimensions: The Well-Separated Pair Decomposition and Its Applications},
  school = {The Johns Hopkins University},
  address = {Baltimore, Maryland},
  year = {1995}
}

@inproceedings{CallahanKosaraju1995Dynamic,
  author    = {Paul B. Callahan and S. Rao Kosaraju},
  title     = {Algorithms for Dynamic Closest Pair and
               {$n$}-Body Potential Fields},
  booktitle = {Proceedings of the Sixth Annual ACM-SIAM
               Symposium on Discrete Algorithms},
  pages     = {263--272},
  year      = {1995},
  url       = {https://dblp.org/rec/conf/soda/CallahanK95}
}

@incollection{MitchellMulzer2017Proximity,
  author = {Joseph S. B. Mitchell and Wolfgang Mulzer},
  title = {Proximity Algorithms},
  booktitle = {Handbook of Discrete and Computational Geometry},
  editor = {Jacob E. Goodman and Joseph O'Rourke and Csaba D. T{\'o}th},
  edition = {3},
  publisher = {CRC Press},
  year = {2017},
  pages = {849--874},
  url = {https://page.mi.fu-berlin.de/mulzer/pubs/proximity.pdf}
}

@book{preparataShamos1985,
  author = {Franco P. Preparata and Michael Ian Shamos},
  title = {Computational Geometry: An Introduction},
  publisher = {Springer},
  year = {1985},
  doi = {10.1007/978-1-4612-1098-6}
}

@misc{Chan2006Minimalist,
  author = {Timothy M. Chan},
  title = {A Minimalist's Implementation of an Approximate Nearest Neighbor Algorithm in Fixed Dimensions},
  year = {2006},
  note = {Manuscript, May 9},
  url = {https://tmc.web.engr.illinois.edu/sss.pdf}
}

@inproceedings{AluruSevilgen1999,
  author = {Srinivas Aluru and Fatih E. Sevilgen},
  title = {Dynamic Compressed Hyperoctrees with Application to the {N}-body Problem},
  booktitle = {Foundations of Software Technology and Theoretical Computer Science},
  series = {Lecture Notes in Computer Science},
  volume = {1738}, pages = {21--33}, year = {1999},
  doi = {10.1007/3-540-46691-6_2}
}

@inproceedings{KhramtcovaLoffler2017,
  author = {Elena Khramtcova and Maarten L{\"o}ffler},
  title = {Dynamic Stabbing Queries with Sub-logarithmic Local Updates for Overlapping Intervals},
  booktitle = {Computer Science -- Theory and Applications (CSR)},
  series = {Lecture Notes in Computer Science}, volume = {10304},
  pages = {176--190}, year = {2017},
  doi = {10.1007/978-3-319-58747-9_17}
}

@misc{HoogLoffler2018Stabbing,
  author = {Ivor van der Hoog and Maarten L{\"o}ffler},
  title = {Approximate Stabbing Queries with Sub-logarithmic Local Replacement},
  year = {2018}, note = {EuroCG 2018 extended abstract, paper 59},
  url = {https://conference.imp.fu-berlin.de/eurocg18/download/paper_59.pdf}
}

@inproceedings{MountPark2010,
  author = {David M. Mount and Eunhui Park},
  title = {A Dynamic Data Structure for Approximate Range Searching},
  booktitle = {Proceedings of the 26th Annual Symposium on Computational Geometry},
  year = {2010}, doi = {10.1145/1810959.1811002}
}

@inproceedings{ParkMount2012,
  author = {Eunhui Park and David M. Mount},
  title = {A Self-adjusting Data Structure for Multidimensional Point Sets},
  booktitle = {Algorithms -- ESA 2012},
  year = {2012}, doi = {10.1007/978-3-642-33090-2_67}
}

@misc{KauerMulzer2019Disks,
  author = {Alexander Kauer and Wolfgang Mulzer},
  title = {Dynamic Disk Connectivity},
  year = {2019},
  note = {35th European Workshop on Computational Geometry (EuroCG), extended abstract, paper 50},
  url = {https://www.eurocg2019.uu.nl/papers/50.pdf}
}

@article{WillardLueker1985,
  author = {Dan E. Willard and George S. Lueker},
  title = {Adding Range Restriction Capability to Dynamic Data Structures},
  journal = {Journal of the ACM},
  volume = {32}, number = {3}, pages = {597--617}, year = {1985},
  doi = {10.1145/3828.3839}
}

@article{Eppstein2008Skip,
  author  = {David Eppstein and Michael T. Goodrich and Jonathan Z. Sun},
  title   = {Skip Quadtrees: Dynamic Data Structures for Multidimensional Point Sets},
  journal = {International Journal of Computational Geometry \& Applications},
  volume  = {18}, number = {1--2}, pages = {131--160}, year = {2008},
  doi     = {10.1142/S0218195908002568}
}

@inproceedings{HartmanisSimon1974,
  author    = {Juris Hartmanis and Janos Simon},
  title     = {On the Power of Multiplication in Random Access Machines},
  booktitle = {15th Annual Symposium on Switching and Automata Theory (SWAT 1974)},
  pages     = {13--23},
  publisher = {IEEE},
  year      = {1974},
  doi       = {10.1109/SWAT.1974.20}
}

@inproceedings{PrattRabinStockmeyer1974,
  author    = {Vaughan R. Pratt and Michael O. Rabin and Larry J. Stockmeyer},
  title     = {A Characterization of the Power of Vector Machines},
  booktitle = {Proceedings of the Sixth Annual ACM Symposium on Theory of Computing (STOC 1974)},
  pages     = {122--134},
  publisher = {ACM},
  year      = {1974},
  doi       = {10.1145/800119.803892}
}

@misc{HarPeled2010Quadtrees,
  author = {Sariel Har-Peled},
  title  = {Quadtrees -- Hierarchical Grids},
  year   = {2010},
  note   = {Lecture notes. A later version is Chapter~2 of \emph{Geometric Approximation Algorithms}, AMS, 2011, doi:10.1090/surv/173/02}
}

\appendix
\newpage
\section{Edge-oracle trees}
\label{app:edge-oracles}

We show how to find the stopping node of the point-location
procedure in Section~\ref{sec:our_quadtree} without following
its entire search path. 

\edgeoracletheorem*

\begin{proof}
Fix a query point $q\in B$, and let $z$ be the node where the
top-down search stops.

\subparagraph*{The edge oracle.}
For an edge from a parent $u$ to a child $v=(C,R)$, the node
$z$ lies in the subtree of $v$ exactly when $q\in R$.
Indeed, all descendants of $v$ have their range contained in~$R$, and all sibling nodes have their ranges disjoint. 
Thus one rectangle-membership test tells us which component of
the tree minus $uv$ contains $z$. This test is
the \emph{edge oracle}.

\subparagraph*{The top tree.}
We maintain a top tree~\cite[Theorem~2.1]{Alstrup2005TopTrees} over the undirected quadtree.
It is a binary tree of height $O(\log n)$ whose leaves are the
quadtree edges. Each node represents a connected subtree,
called a \emph{cluster}. For any node in the top tree, its child clusters meet at one vertex,
and their union is the parent cluster.

A cluster has at most two boundary vertices, meaning vertices
incident to an edge outside the cluster. At each boundary vertex, we
store the at move five incident edges belonging to the cluster. We also store the vertex shared by two child clusters.

\subparagraph*{Searching.}
Start at the top-tree root: this cluster contains all quadtree nodes and thus also the node $z$ that is the answer to our query.
Suppose the current cluster has two children meeting at $x$, which is a quadtree node. 
Test the quadtree edges incident to $x$. If every oracle places $z$
on the side containing $x$, then $x = z$ and so return $x$.
Otherwise, exactly one edge $e$ incident to $x$ separates $x$ from $z$.
This is the first edge on the path from $x$ to $z$.
This edge is contained in exactly one of its two children, and that child is the child that stores the path from $x$ to $z$ (since the path cannot enter the other
child without revisiting $x$, their only shared vertex).
We continue the search in that child.
Each step takes constant time and descends one top-tree level. At a top-tree leaf, the cluster consists of one edge.
Since the cluster contains $z$, one of its endpoints is $z$,
and one oracle test identifies which endpoint to return.
Each step takes constant time and descends one top-tree level,
giving $O(\log n)$ worst-case query time.

\subparagraph*{Updates and space.}
A point update first changes the quadtree $T$.
By Lemma~\ref{lem:insertion} and the rebuilding schedule of
Section~\ref{sec:deletions}, this requires only $O(1)$ local
changes after point location.
We then update the top tree to reflect each inserted or deleted
edge of $T$.

For each such edge change, the top tree inserts or removes the
corresponding leaf and reorganises its hierarchy using
$O(\log n)$ cluster splits and joins.
A split separates a cluster into its two children.
A join creates a parent cluster from two parent-less clusters that share exactly one vertex.
For each boundary vertex of the parent, we combine the children's
lists of incident edges at that vertex.
Since these lists have constant size, maintaining them takes
constant time per split or join.
Thus each edge change in $T$ takes $O(\log n)$ time. 

A change to a quadtree range requires no change to the top tree,
since oracle tests read the current ranges directly.
Including point location, the total time per point update is
therefore $O(\log n)$ in the worst case.
The top tree uses $O(n)$ space. 
\end{proof}

\newpage

\section{A well-separated pair decomposition}
\label{app:wspd}

For two point sets $A$ and $B$, we write $\dist(A,B)$ for the minimum distance between
a point in $A$ and a point in $B$, and $\diam(A)$ for the maximum distance between two points in $A$.

\begin{definition}
\label{def:WSPD}
    An $s$-WSPD of a point set $P$ is a collection of pairs $(A,B)$ of disjoint
non-empty subsets of $P$ such that every unordered pair of distinct
points of $P$ lies across exactly one pair $(A,B)$, and each
$(A,B)$ satisfies $\dist(A,B)\geq s\max\{\diam(A),\diam(B)\}$.
\end{definition}

\begin{theorem}\label{thm:wspd}
Fix a constant $M$. Let $T$ be an aligned quadtree storing a set $P$ of $n$ points in the plane with $O(n)$
nodes and the packing property with that fixed constant $M$. For any $s\geq 1$, we can compute in
$O(s^2n)$ time an $s$-WSPD consisting of $O(s^2n)$ pairs of nodes
of $T$ (each node represents the points stored in the leaves of its subtree). 
\end{theorem}

\begin{proof}
We say a node is \emph{occupied} if there is at least one point stored in its subtree. For each occupied node $v$, let $P_v$ be the points in its subtree and
$B_v$ be their smallest closed axis-aligned bounding box. These bounding boxes can be computed in linear total time via a bottom-up procedure, and each $B_v$ lies in the range of $v$.

\subparagraph{Computing the pairs.}
For each unordered pair of occupied siblings $u,v$ in $T$ we call the function
$\textsc{FindPairs}(u,v)$, which is is the following recursive function:
\begin{itemize}[nolistsep]
    \item If $\dist(B_u,B_v)\geq s\max\{\diam(B_u),\diam(B_v)\}$,
    output $(u,v)$ and return.
    \item Else, assume that
    $\diam(B_u)\geq\diam(B_v)$ (otherwise,  swap the roles of $u$ and $v$). Call $\textsc{FindPairs}(w,v)$ for
    each occupied child $w$ of $u$. We call this \emph{splitting} $u$.
\end{itemize}
We show that each pair $\{p,q\}$ of distinct points appears in a pair $(u,v)$ with $p \in P_u$ and $q \in P_v$ (or the other way around) exactly
once. The lowest common ancestor $v(p, q)$ of
the leaves storing $p$ and $q$ has distinct children $u$ and $v$ whose ranges contain $p$ and $q$, respectively. Hence $\textsc{FindPairs}(u,v)$ is the unique
highest level call covering $\{p,q\}$. Whenever such a call splits~$u$,
exactly one child of $u$ contains $p$, so exactly one recursive call
still covers $\{p,q\}$. Splitting $v$ is symmetric. Each split
descends the tree, and two occupied leaves pass the separation test
because their boxes have diameter zero. Thus exactly one output pair
covers $\{p,q\}$, and it is $s$-well-separated by the test.

\subparagraph{Running time.}
Fix a node $u$ that is split. Let $\Pi$ consist of its \emph{partners},
the nodes $v$ against which a call splits $u$. Let $d:= \diam(B_u)$. Since the separation
test fails and $B_u$ has the larger diameter, we have that $d>0$. We use $d$ and $\Pi$ to invoke the packing property (Definition~\ref{def:packing}).

Every partner's range meets the disk $D$ centred at the centre $x$
of $B_u$ with radius $r=(s+\tfrac12)d$. Indeed, the failed test
gives $\dist(B_u,B_v)<sd$. Since every point of $B_u$ lies within
$d/2$ of $x$, the disk meets $B_v$ and hence the range of $v$.

Next, we note that partners from distinct calls have disjoint point sets. Indeed, choose for the sake of contradiction a shared point $q$ and any $p\in P_u$. Both calls split $u$
while covering $\{p,q\}$, yet after the first split this pair
continues in the child of $u$ containing $p$. Thus no partner occurs
twice and no node in $\Pi$ is an ancestor of another.

We finally show that the lowest common ancestor $a$ of any two distinct
partners $v,v'$ has cell side length at least $d/\sqrt{2}$.
If $a$ is also an ancestor of $u$, then $B_u\subseteq B_a$ and so this follows.
Otherwise, let $b$ be the lowest common ancestor of $u$ and $v$.
Since $v$ is a descendant of $a$ and $u$ is not, $b$ an ancestor of $a$. The earlier call that lead to us splitting $u$ is on the children of $b$
leading to $u$ and $v$, respectively. The recursive path to $v$ passes through
$a$. Since $v$ is a proper descendant
of $a$, the recursion must split $a$.
Suppose the call that splits $a$ pairs it with a node $w$. This call
eventually leads to the call on $u$ and $v$, with $v$ reached from
$a$ and $u$ reached from $w$. Hence $B_u\subseteq B_w$.
Since the algorithm splits $a$, we have
$\diam(B_a)\geq\diam(B_w)\geq\diam(B_u)=d$.  The cell of $a$ contains $B_a$, so its diagonal is
at least $d$ and its side length is at least $d/\sqrt{2}$.

By the packing property (Definition~\ref{def:packing}), applied to
$\Pi$ and $D$ with $r=(s+\tfrac12)d$ and $\rho=d/\sqrt{2}$, we have
$|\Pi|\leq M(1+r/\rho)^2=O(s^2)$. Since each partner occurs only
once, at most $O(s^2)$ calls split $u$.
Each node has at most six unordered pairs of occupied children, so there are $O(n)$ initial calls. Each call that splits a node makes at most four recursive calls. Thus there are $O(s^2n)$ calls in total. Each call takes constant time and outputs at most one pair.
\end{proof}

\section{Approximate nearest neighbours}\label{app:ann-recovery}
In this section, we show how to adapt our aligned quadtree such that approximate nearest neighbour queries can be answered efficiently on a Real-RAM.

\begin{theorem}\label{thm:ann-recovery}
Let $P$ be a dynamic set of $n$ points in a fixed bounding square.
There is a data structure that uses
$O(n\log^3 n)$ space and supports updates in $O(\log^5 n)$ 
amortised time such that given a query point $q$ and $0<\varepsilon\leq1$ it can find a
$(1+\varepsilon)$-approximate nearest neighbour of $q$ in $P$ in
$O(\log^4 n+\varepsilon^{-2}\log n)$ time. If $P$ is empty, we report this immediately.
\end{theorem}

\begin{proof}
Let $P$ be non-empty and let $p^*$ be
a nearest neighbour of $q$. The data structure we construct storing $P$ consists of our aligned quadtree (and its edge-oracle tree), together with two range trees. The first range trees is used to find some search radius $r$, and the second is used to find a subset of quadtree nodes that we call \emph{selected} nodes using this radius. We finally show that there is a point in one of the selected quadtree nodes that is an approximate nearest neighbour of $q$, and we can thus report such a point efficiently. 

We first show how to construct a range tree that we can use to compute the radius $r$
telling us how far to search around $q$ for (an approximation of) $p^*$.

\subparagraph{Choosing the search radius.}
Let $r=\min_{p\in P}\|p-q\|_\infty$. At least one input point is within
Euclidean distance $\sqrt{2}r$ of $q$, and every input point is at
Euclidean distance at least $r$ from $q$. Hence
$r\leq\|p^*-q\|_2\leq\sqrt{2}r$. In particular, it suffices to find
a point within distance $\varepsilon r$ of $p^*$.

To compute $r$, we maintain a separate two-dimensional range tree
that stores $(x+y,x-y)$ for each point $(x,y)\in P$. We write
$(u_p,v_p)$ for the transformed coordinates of a point $p$. For each $p\in P$, the $L_1$ distance
between the transformed points of $p$ and $q$ is
$|u_p-u_q|+|v_p-v_q|=2\|p-q\|_\infty$.
We use $4$ range queries to find the smallest such distance.

The first query asks for a point $p$ minimising $u_p+v_p$ among all
points with $(u_p,v_p)\in[u_q,\infty)\times[v_q,\infty)$.
Indeed, for every point in this range, the distance is
$(u_p-u_q)+(v_p-v_q)=(u_p+v_p)-(u_q+v_q)$, and $u_q+v_q$ is fixed.
The other three queries follow by the same calculation.
For $u_p\leq u_q$ and $v_p\geq v_q$, we minimise $-u_p+v_p$.
For $u_p\geq u_q$ and $v_p\leq v_q$, we minimise $u_p-v_p$.
For $u_p\leq u_q$ and $v_p\leq v_q$, we minimise $-u_p-v_p$.

For each node $\mu$ of the range tree, we store four
pointers. Let $S_\mu \subseteq P$ be the set of points stored in $\mu$. The first pointer points to a point $p\in S_\mu$ minimising $u_p+v_p$. The other three point to points
in $S_\mu$ with smallest $-u_p+v_p$, $u_p-v_p$, and $-u_p-v_p$,
respectively. To answer the first query, the range tree decomposes
its query range into $O(\log^2 n)$ canonical subsets of stored points.
We follow the first pointer for each of these subsets and return the
point with the smallest $u_p+v_p$ among them. The other three queries
are answered in the same way.
We take the smallest $L_1$ distance from $(u_q,v_q)$ to any point
returned by the four queries and divide it by two to obtain $r$. If $r=0$, we return the point attaining this minimum.

\subparagraph{Selecting quadtree nodes.}
Set $\rho=\varepsilon r/2$.
For each internal node $v$ of the quadtree, let $h(v)$ be the length of the longest
side of its range. Set $h(v)=0$ at a leaf, since it stores at most
one point.
We call a quadtree node \emph{selected} if it is the first node on its path from
the root with $h(v)\leq\rho$. Every input point belongs to exactly
one selected node. All points in a selected node have pairwise distance at most $\sqrt{2}\rho\leq\varepsilon r$. In particular, any point inside the selected node whose range contains $p^*$ gives the required approximation.
 We next show how to report only the
selected nodes near $q$, including the one which contains $p^*$.

\subparagraph{Reporting selected nodes.}
For each internal node or occupied leaf $v$ other than the root, we store the four-dimensional point $(c_x(v),c_y(v),h(v),h(\operatorname{parent}(v)))$ in a range tree.
Here, $c(v)$ is the lower-left corner of the range of $v$ or, if $v$ is a leaf, its stored point, and $c_x(v)$ and $c_y(v)$ denote its $x$- and $y$-coordinate, respectively.
We use the range tree to report all nodes with $h(v)\leq\rho<h(\operatorname{parent}(v))$ and $\|c(v)-q\|_\infty\leq 2r+\rho$, testing the root separately.
This takes $O(\log^4 n + k)$ time, where $k$ is the number of reported nodes. The output includes the selected node containing $p^*$, since its $c(v)$ is within $\rho$ of $p^*$ in each coordinate and $\|p^*-q\|_2\leq 2r$.

We next bound the number of reported nodes.
Let $\Pi$ be the set of reported nodes and let $D$ be the disk centred
at $q$ of radius $R=\sqrt{2}(2r+\rho)$. We verify the three conditions
of Definition~\ref{def:packing}. First, for each $v\in\Pi$, the disk
$D$ contains $c(v)$, so the range of $v$ intersects or touches $D$.
Second, no node in $\Pi$ is an ancestor of another, since only the
first node on each path with $h(v)\leq\rho$ is chosen. Third, the
lowest common ancestor $u$ of any two nodes in $\Pi$ precedes both
chosen nodes on their paths. Thus $h(u)>\rho$. Since the cell of $u$
contains its range, its side length is greater than $\rho$.
By the packing property (Definition~\ref{def:packing}), applied to
$\Pi$ and $D$ with radius $R=\sqrt{2}(2r+\rho)$ and threshold
$\rho=\varepsilon r/2$, we have
$|\Pi|\leq M(1+R/\rho)^2=O(\varepsilon^{-2})$.

\subparagraph{Returning the output.}
For each $v\in\Pi$, we choose an arbitrary point in its subtree if it contains one. We can find such a point using the edge-oracle tree. On the edge entering each occupied leaf, we store a pointer to its point. In each cluster of the top tree (Appendix~\ref{app:edge-oracles}), we store one such pointer if the cluster contains an edge with a pointer.

If $v$ is a leaf, we take its stored point. If $v$ is internal, we cut the edge to its parent, read the pointer at the root of the top tree for the component containing $v$, and restore the edge. If $v$ is itself the root, we simply read the pointer for the whole tree. A null pointer means that the subtree contains no point, in which case we skip $v$. This takes logarithmic time per reported node. We return the chosen point closest to $q$ in Euclidean distance.
Consider the node $v^* \in \Pi$ that contains~$p^*$ and its chosen point  $p'$. 
The point returned is at least as close to $q$ as $p'$, so its distance to $q$ is at most $\|p^*-q\|_2+\varepsilon r\leq(1+\varepsilon)\|p^*-q\|_2$.

\subparagraph{Updates and space.}
Each local quadtree update changes $O(1)$ nodes and child links.
Since each node has constant degree, only $O(1)$ range-tree points
change. When a leaf becomes occupied or empty, we update the pointer on its incoming edge, if it has one, and in the $O(\log n)$ clusters containing that edge. At each join, we take a pointer from either child that has one, and a null pointer if neither does. This takes constant time per cluster and $O(n)$ space in total. During the global rebuilding of
Section~\ref{sec:deletions}, we maintain all structures for both
the current tree and its replacement. The dynamic range
trees~\cite{WillardLueker1985} use $O(n\log^3 n)$ space and have
$O(\log^5 n)$ amortised update time.
\end{proof}

\section{Errors in Callahan's dynamic construction}
\label{app:callahan}

Callahan~\cite[Chapter~6]{Callahan1995Thesis} claims a dynamic explicit well-separated pair decomposition (WSPD) construction for a point set in $\mathbb{R}^d$ based on a tree they call a fair-split tree.
They claim that by inserting dummy points into this fair-split tree, one can maintain a linear-size WSPD for a dynamic point set under the Euclidean metric.
Har-Peled~\cite[p.~43]{HarPeled2011} comments on this feat when citing it:
``This is quite surprising when one considers that, in the worst
case, a point might participate in a linear number of pairs''.
In other words, it is well-known that a single point can participate in a linear number of WSPD pairs, making it surprising that an explicit construction could be maintained with only polylogarithmic update time.
We believe that this result by Callahan is incorrect. 
We show that it is incorrect as written, as we give counterexamples to three critical constructions in the paper. 
Unless otherwise stated we consider the case of points in the plane.
For the sake of verification, a copy of the thesis can be found at this 
\href{https://anonymous.4open.science/r/PublicPapers-F74D/}{anonymous repository}.

\subparagraph{The insertion rule can be forced to fail.}
The first error we found does not pertain to the WSPD but rather to the stand-alone fair split tree data structure.
Callahan's insertion rule~\cite[pp.~117--121]{Callahan1995Thesis}
uses a binary tree where each node $v$ stores a dimensional rectangle $R(v)$ containing the points stored in the subtree rooted at $v$. A leaf storing a point $z$ has rectangle
$\{z\}$. For a rectangle, write $l_{\max}(R)$ for the longest side of $R$.
A fair cut splits a rectangle $R$ in two  by an axis-parallel line at distance
at least $l_{\max}(R)/3$ from both parallel sides. At each internal node $v$, such a cut partitions $R(v)$
into two cut rectangles $R_1$ and $R_2$. Each child $w_i$ stores
a rectangle $R(w_i)\subseteq R_i$, which may be strictly
smaller than $R_i$. Thus the children's stored rectangles need not
partition $R(v)$.

\begin{figure}
\centering
\includegraphics{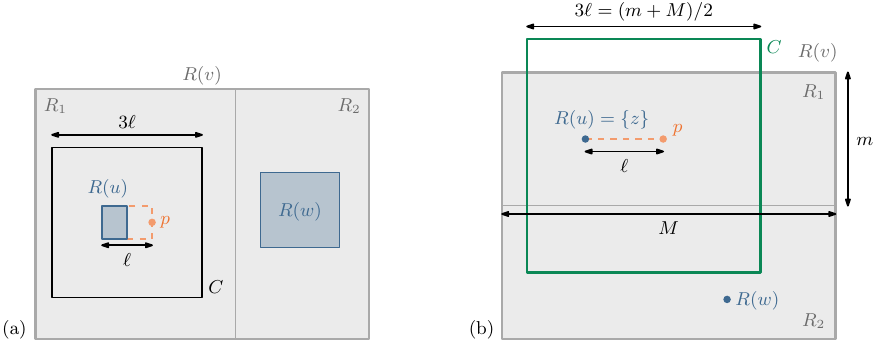}
\caption{(a) The existing cut partitions $R(v)$ into $R_1$ and $R_2$.
The new point $p$ lies in $R_1\setminus R(u)$.
Callahan inserts a new node above $u$ and requires
a square $C\subseteq R_1$ of side length $3\ell$ containing $R(u)$,
where $\ell$ is the longest side of the bounding rectangle of $R(u)\cup\{p\}$ (dashed orange).
(b) The node $v$ has two leaf children $u,w$, whose stored rectangles
are the blue singleton points in $R_1,R_2$, respectively.
For $\ell=(m+M)/6$, the required square $C$ with side length
$3\ell=(m+M)/2>m$ and cannot fit inside $R_1$.}
\label{fig:callahan-insertion}
\end{figure}

To insert a new point $p$ lying in the root's rectangle,
Callahan's algorithm~\cite[p.~118]{Callahan1995Thesis} finds the
deepest node $v$ whose rectangle contains $p$, see Figure~\ref{fig:callahan-insertion}(a).
Since each leaf stores a singleton rectangle, $v$ is not a leaf.
Let $u$ and $w$ be its children. By the choice of $v$,
$p\notin R(u)$ and $p\notin R(w)$.
The existing fair cut at $v$ partitions $R(v)$ into two cut
rectangles $R_1$ and $R_2$, with $R(u)\subseteq R_1$ and
$R(w)\subseteq R_2$.
Suppose $p\in R_1$. Callahan's algorithm inserts a new node
between $v$ and $u$, with $u$ and a new leaf storing $p$ as its
children. To construct the rectangle of this new node, it sets
$\ell$ to the longest side of the bounding rectangle of
$R(u)\cup\{p\}$. If $\ell<l_{\max}(R_1)/3$, the algorithm requires
a square $C$ contained in $R_1$ of side length $3\ell$
containing $R(u)$~\cite[p.~120]{Callahan1995Thesis}.
We show that this square need not exist.

Every fair-split tree with at least two leaves has an internal node
$v$ whose children $u$ and $w$ are both leaves. Let $R_1$ be the
cut rectangle at $v$ containing $R(u)$, and suppose that $R_1$
has side lengths $m<M$. Let $z$ be the point stored at $u$, so
$R(u)=\{z\}$; see Figure~\ref{fig:callahan-insertion}(b).
Choose a new point $p$ at distance $\ell=(m+M)/6$ from $z$ along
the longer axis towards the farther side of $R_1$.
Then $p\in R_1$, since $\ell<M/2$.
Furthermore, $p\notin R(u)$ and $p\notin R(w)$, since $R(w)\subseteq R_2$.
Thus $v$ is the deepest node whose rectangle contains $p$.
The bounding rectangle of $R(u)\cup\{p\}=\{z,p\}$ has longest
side $\ell<M/3$, and thus Callahan's insertion rule requires
a square $C\subseteq R_1$ of side length $3\ell$ containing $R(u)$.
This is impossible since $3\ell=(m+M)/2>m$.
Thus the insertion rule is not defined for every valid
fair-split tree and insertion point. We briefly note, that this problem is essentially a version of the alignment problem for quadtrees. Our paper resolves this precisely by allowing a child of a node to 'stick out' of its parent. This property seems to be necessary, although we give no formal proof for that statement. But, as one can see in this paper, allowing children to stick out of the parent  out generates other problems including the fact that the packing property of quadtrees is no longer automatically guaranteed.

\subparagraph{The dummy-point loop adds unbounded many points.}
Even if the fair-split tree were well-defined and well-maintained, the specific algorithm that maintains a WSPD using this tree has additional flaws.
Callahan~\cite[Invariant~6.3.2 and Lemma~6.3.3, pp.~124--125]{Callahan1995Thesis}
adds dummy points to the tree.
The idea of these dummy points is to prevent that a single insertion requires linearly many updates to the WSPD: to 'pre-emptively' and 'gradually' add WSPD pairs with these dummy points so that if we add a new point $p$, the tree already had ancestors of $p$ in the necessary WSPD pairs before the insertion.
The thesis claims that these dummy points add only a constant-factor overhead.

Each node $v$ stores a rectangle $R(v)$ containing
the real and dummy points in its subtree. Dummy points are not counted as real points.
The claimed bound is $O(1)$ dummy points for each well-separated
pair of nodes whose subtrees both contain real points.
Callahan groups distances into \emph{distance classes}%
~\cite[p.~123]{Callahan1995Thesis}. We write $\eta$ for his
distance-class parameter to distinguish it from other approximation
parameters in this paper. This distance-class parameter $\eta$ is distinct from the WSPD
separation parameter $s>0$ in Definition~\ref{def:WSPD}.
Callahan tests separation using the stored rectangles: they must
fit in two balls of equal radius $\rho$ with distance at least
$2s\rho$ between the balls~\cite[pp.~24, 124]{Callahan1995Thesis}.
Here we write his separation constant as $2s$, so that this condition
implies the separation inequality of Definition~\ref{def:WSPD}
for the point sets below the two nodes.
In particular, overlapping rectangles cannot
form such a pair as they have distance zero.

For fixed $0<\eta<1$ and a scale $r>0$,
the distance class is defined as
\[
\mathrm{DC}_r=[(1-\eta)r,(1+\eta)r).
\]
Thus a distance belongs to $\mathrm{DC}_r$ if it is within a
multiplicative factor $1\pm\eta$ of $r$. For distinct real points $y,z$, Callahan's
Invariant~6.3.2 requires for every point $x$ with
\[
\|y-x\|_2\in\mathrm{DC}_{\|y-z\|_2/(1-\eta^2)},
\qquad\text{equivalently}\qquad
\frac{\|y-z\|_2}{1+\eta}
\leq \|y-x\|_2 < \frac{\|y-z\|_2}{1-\eta},
\]
that there is well-separated pair of nodes $a,b$ with $y$
in the substree of $a$ and $x\in R(b)$.
If this requirement fails, Callahan inserts an
uncovered point $x$ as a dummy point and repeats, allowing $x$ to be
``selected arbitrarily from any space left uncovered''%
~\cite[p.~125]{Callahan1995Thesis}.

\begin{figure}[tb]
\centering
\includegraphics{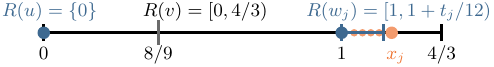}
\caption{Our construction to insert an unbounded number of dummy points. The only real points (blue) that are stored are $0$ and $1$, while we can insert many dummy points (orange). The insertion of $x_j$ incurs the creation of a new node in the split tree, because it lies outside $R(w_j)$.}
\label{fig:callahan-dummy}
\end{figure}

We provide an example where this procedure can lead to inserting an unbounded number of dummy points in the 1-dimensional case, i.e., for points on a line.
Consider the tree whose root $v$ stores the interval
$R(v)=[0,4/3)$ that is cut at $8/9$. Its two children are
leaves $u,w_1$ with $R(u)=\{0\}$ and $R(w_1)=\{1\}$.
This is a valid fair-split tree.
The only real points are $0$ and $1$.
Fix any positive integer $N$, and choose $t_1>0$ so that
$t_j=16^{j-1}t_1$ satisfies $t_N<\min\{\eta,1/100\}$.
We show that the loop can insert all $N$ dummy points $x_j=1+t_j$ in order,
see Figure~\ref{fig:callahan-dummy}.

Before inserting $x_j$, let $w_j$ be the right child of $v$.
Inductively, $R(w_j)=[1,1+t_j/12)$ for $j>1$, with
$R(w_1)=\{1\}$ initially. Thus $x_j$ lies in $R(v)$ but outside
both rectangles of $v$'s children. The point distance $||0 - x_j||_2 = 1 + t_j < 1 + \eta$ is in the distance class $\mathrm{DC}_{||0-1||_2/(1-\eta^2)}$, and so the point $x_j$ is required to be covered by a well-separated pair of nodes $a,b$ with the real point $0$ in the subtree of $a$ and $x \in R(b)$.
Only $R(v)$ contains $x_j$, and it
overlaps every node rectangle containing $0$. Hence no
such well-separated pair $a,b$ exists.

To insert $x_j$, Callahan's rule~\cite[pp.~120--121]{Callahan1995Thesis}
uses the right cut interval $[8/9,4/3)$ at~$v$.
The bounding interval of $R(w_j)\cup\{x_j\}$ has length $t_j<|[8/9,4/3)|/3$, and thus we require an interval $C$ that is contained in $[8/9,4/3)$, has side length $3 t_j$, and contains $R(w_j)$.
The rule permits choosing $C=[1,1+3t_j)$.
Callahan then iteratively decreases the size of $C$ towards $R(w_j)$, using an endpoint
of $R(w_j)$ if it lies in $C$'s middle third,
and otherwise removing the outer third of $C$ farthest from $R(w_j)$.
Here three successive decreases remove the right third, giving lengths
$3t_j,2t_j,4t_j/3,8t_j/9$. The third decrease is the first to exclude
$x_j$, so Callahan sets the rectangle of
the new node $w_{j+1}$, whose children are $w_j$ and the new leaf
storing $x_j$, to be:
\[
 R(w_{j+1})=[1,1+4t_j/3)=[1,1+t_{j+1}/12),
\]
which proves the induction. Since $N$ is arbitrary, there is no
bound on the number of dummy points this procedure adds even though the number of real points is constant. 

\subparagraph{The required set of radii need not be finite.}
The final error concerns the algorithm that maintains directions
on the WSPD pairs. Callahan assigns a direction to each pair to
control the work required by queries and updates. These directions
depend on weights assigned to the points, which the algorithm may
change. To maintain the directions after a weight changes,
Callahan requires a small set of radii satisfying a condition on
these weights. We show that even a fair split tree with two leaves can require
infinitely many radii.

Callahan~\cite[pp.~129--130]{Callahan1995Thesis} states each leaf $v$ in the fair split tree $T$ has an integer weight $w(v)$ between $0$ and $|T|$, where $|T|$ is the
number of nodes. Each internal node has weight equal to the sum of
the weights of the leaves in its subtree. Zero weights are required by Callahan's insertion and deletion
procedures: the leaf storing the point being inserted or deleted
must have weight zero~\cite[Section~6.6, p.~143]{Callahan1995Thesis}.
To measure the weight near a point, Callahan defines the
\emph{$\delta$-rendering} of a square $C$%
~\cite[pp.~129--130]{Callahan1995Thesis}.
Here $\delta>0$ is a fixed parameter.
Write $l(C)$ for the side length of $C$, $l_{\max}(R)$ for the
longest side of a rectangle $R$, and $V(T)$ for the set of nodes
of $T$. Callahan defines the $\delta$-rendering as the following set:
\[
\rho_\delta(C):=
\left\{
v\in V(T)\setminus\{\operatorname{root}(T)\}:
\begin{array}{l}
R(v)\cap C\neq\emptyset,\\
l_{\max}(R(v))\leq\delta l(C)
<l_{\max}(R(\operatorname{parent}(v)))
\end{array}
\right\}.
\]
Thus $\rho_\delta(C)$ is a set of nodes whose rectangles intersect
$C$ and are small enough, but whose parents'
rectangles are not.

Suppose the weight of a leaf $u$ storing a point $x$ changes.
Callahan defines a weight function $w_{\min}$ on all nodes:
$w_{\min}(u)$ is the smaller of the old and new weights of $u$;
for every other leaf $v$, $w_{\min}(v)=w(v)$; and for every
internal node $v$, $w_{\min}(v)$ is the sum of the
$w_{\min}$ weights of the leaves in its subtree. For a set $S$ of nodes, write
$w_{\min}(S)=\sum_{v\in S}w_{\min}(v)$.

For each radius $r>0$, let $C_r$ be the square centred at $x$
with side length $2r$. For a fixed constant $\beta>1$, the square
$C_{\beta r}$ has the same centre and side length $2\beta r$.
Callahan compares the weights of the renderings of these two
squares $C_r$ and $C_{\beta r}$. For another fixed constant $\kappa>0$, whenever
\[
 w_{\min}(\rho_\delta(C_{\beta r}))
 \geq \frac{\kappa+1}{\kappa}\,
       w_{\min}(\rho_\delta(C_r)),
\]
his algorithm requires a radius $\widehat r\in(r/\beta,r]$
from a set $\mathcal R_x$.
The same set $\mathcal R_x$ must supply such a radius
for every $r>0$ satisfying the displayed inequality.
Lemma~6.5.9 claims that such a set exists
``of size $O(\log |T|)$''~\cite[p.~137]{Callahan1995Thesis}.
However, consider a one-dimensional fair-split tree $T$ with exactly three
nodes: a root $v$ with $R(v)=[0,2)$ and two leaf children $u_0,u_1$. The two cut intervals of $R_v$ are
$R_1=[0,1)$ and $R_2=[1,2)$.
The leaf $u_0$ stores the point $0$, with $R(u_0)=\{0\}$,
and the leaf $u_1$ stores the point $1$, with $R(u_1)=\{1\}$.

Initially, $w(u_0)=w(u_1)=1$. Before deleting the point $0$,
Callahan's algorithm changes $w(u_0)$ from $1$ to $0$.
Consider this weight update, while both leaves are still in the
tree. The point being updated is $x=0$, and
\[
w_{\min}(u_0)=0,\qquad
w_{\min}(u_1)=1,\qquad
w_{\min}(v)=1.
\]

Fix any $r>0$ small enough that $\beta r<1$ and
$\delta\beta r<1$. The intervals $C_r$ and $C_{\beta r}$ centred at $x=0$ are
$C_r=[-r,r]$ and $C_{\beta r}=[-\beta r,\beta r]$.
Since $r<\beta r<1$, neither $C_r=[-r,r]$ nor
$C_{\beta r}=[-\beta r,\beta r]$ contains the point $1$.
Thus $R(u_1)=\{1\}$ intersects neither interval, so
$u_1\notin\rho_\delta(C_r)$ and
$u_1\notin\rho_\delta(C_{\beta r})$.

The node $u_0$ belongs to both $\rho_\delta(C_r)$ and
$\rho_\delta(C_{\beta r})$. Indeed, its rectangle $R(u_0)=\{0\}$
intersects both $C_r=[-r,r]$ and
$C_{\beta r}=[-\beta r,\beta r]$.
Moreover, $R(u_0)$ has length zero, and its parent $v$ has
$R(v)=[0,2)$, which has length $2$.
For $C_r$, the size condition in the definition of the rendering is
\[
l_{\max}(R(u_0))=0
\leq\delta l(C_r)=2\delta r
<2=l_{\max}(R(v)).
\]
For $C_{\beta r}$, the same condition holds because
$
l_{\max}(R(u_0))=0
\leq\delta l(C_{\beta r})=2\delta\beta r
<2=l_{\max}(R(v))$.
Thus $u_0$ satisfies both the intersection and size conditions
for each rendering.

Suppose that $\mathcal R_0$ is finite.
For every $r>0$ with $\beta r<1$ and $\delta\beta r<1$,
the preceding argument shows that
$\rho_\delta(C_r)=\rho_\delta(C_{\beta r})=\{u_0\}$,
since $u_0,u_1$ are the only non-root nodes.
Both renderings therefore have weight $w_{\min}(u_0)=0$.
Callahan's inequality reduces to
$0\geq((\kappa+1)/\kappa)\cdot0$, so $\mathcal R_0$ must
contain a radius in $(r/\beta,r]$ for every such $r$.
In particular, $\mathcal R_0$ must contain a positive radius.
Let $r_*$ be its smallest positive radius, and choose $r>0$
with $r<r_*$, $\beta r<1$, and $\delta\beta r<1$.
Callahan's inequality holds, but $\mathcal R_0$ contains no
radius in $(r/\beta,r]$, since every number in this interval
is positive and smaller than $r_*$.
Thus no finite set $\mathcal R_0$ satisfies Lemma~6.5.9,
even for this tree with two leaves.

\section{Packing in dynamic quadtree applications}
\label{app:literature}

This section presents a survey of papers that, either implicitly or explicitly, use a dynamic or insertion-only quadtree.  In particular, these results rely, in one way or another, on a packing argument of quadtree cells in a dynamic or insertion-only quadtree. 
For each paper that we discuss, we record its model of computations and the relevant assumptions of that model of computation.
We then state their claimed result. 
We then very briefly describe how it uses quadtrees, whether the result transfers to our quadtree, and how
it defines compressed cells.
We discuss Callahan's dynamic fair-split-tree and WSPD construction
separately in Appendix~\ref{app:callahan}.
These were collected by LLMs, but all claims have been verified by humans, several times. We assume full responsibility of the correctness of these claims.

A result \emph{transfers} if our quadtree supports the same functionality
for arbitrary real-valued input in the plane, with polylogarithmic overhead
in the number of objects for fixed~$\varepsilon$. We state larger slowdowns
explicitly. ``Not established'' means that we have no proof of a transfer.

Some papers already assume bounded spread to support real input and keep
their own dependence on the spread.

The packing bullets say whether an upper bound on nearby cells suffices for
the paper, or whether it needs more, such as containment in a shifted cell
of comparable size, smoothness, or a bound near a thin boundary.
The six subsections follow the six approaches listed in the introduction.

\subsection{Finite-coordinate implementations}
\label{app:literature-wordram}
These results restrict coordinate representations or the universe. Some store a canonical tree, others implement its associated orders or grids. 
\subparagraph{Closest-Point Problems Simplified on the RAM~\cite{Chan2002RAM}}
\begin{itemize}
\item \textbf{Model and assumptions.} They assume positive integer coordinates bounded by $U=2^w$ and compare points in shuffle order, obtained by interleaving coordinate bits. Bitwise exclusive-or is the only operation they need beyond arithmetic and comparisons.

\item \textbf{Stated result.} They give an algorithm for closest pair in $O(n\log n)$ time, and a dynamic version in which the $d+1$ sorted lists are kept in standard search trees with logarithmic update time. For $(1+\varepsilon)$-approximate nearest neighbours they give $O(\varepsilon^{-d}\log n)$ query time with $O(\log n)$ update time, and $O(\log n+\varepsilon^{-d})$ query time after a further modification.

\item \textbf{Usage of packing properties: Shifted-cell containment.} The correctness argument uses two facts: points in a quadtree box are consecutive in shuffle order, and each pair of points is contained in a comparable-size box after one of the prescribed shifts. These are stronger requirements than an upper bound on nearby cells.

\item \textbf{Result transfer: Not established.} Our packing property does not supply the fixed shifted orders or the comparable-size containment guarantee.

\item \textbf{Compressed definition: Implicit canonical hierarchy.} They explicitly avoid maintaining a quadtree and instead compare shuffle orders. 
\end{itemize}

\subparagraph{A Minimalist's Implementation of an Approximate Nearest Neighbor Algorithm in Fixed Dimensions~\cite{Chan2006Minimalist}}
\begin{itemize}
\item \textbf{Model and assumptions.} In Section~1 they state that the algorithm does not work on the Real-RAM: each coordinate is an integer fitting in one word, and exclusive-or and most-significant-bit operations on words are available.

\item \textbf{Stated result.} For fixed $\varepsilon$ and $d$: $O(n\log n)$ preprocessing, no space beyond the input array, and $O(\log n)$ expected query time (Section~1). The expectation is over the algorithm's random shift and assumes queries that are independent of it. The dynamic version (Section~3) keeps the points in a balanced search tree and supports insertions and deletions in $O(\log n)$ time.

\item \textbf{Usage of packing properties: Shifted-cell containment.} In Section~2, they use canonical boxes and the probability that nearby points share a small box after shifting. 

\item \textbf{Result transfer: Yes for planar approximate nearest-neighbour functionality, with polylogarithmic overhead.} Theorem~\ref{thm:ann-recovery} gives $O(\log^4n+\varepsilon^{-2}\log n)$ queries and $O(\log^5n)$ amortised updates using $O(n\log^3n)$ space. It replaces the search algorithm. 

\item \textbf{Compressed definition: Implicit canonical hierarchy.} In Section~2 they compare points in the order obtained by interleaving their coordinate bits. 
\end{itemize}

\subparagraph{A Succinct, Dynamic Data Structure for Proximity Queries on Point Sets~\cite{VenkatMount2014}}
\begin{itemize}
\item \textbf{Model and assumptions.} In Section~1 they state that they use the Word-RAM, with coordinates represented by $b$ bits.

\item \textbf{Stated result.} Let $m$ be the number of nodes of the uncompressed quadtree, which can exceed the number $n$ of points, and let $\Phi$ be the aspect ratio of the point set. Theorem~1 (static) stores the tree in $(d+2)m+o(m)$ bits and answers $\varepsilon$-approximate nearest-neighbour queries in $O(\varepsilon^{-d}\log\Phi\,\lg^2\lg m)$ time and $\varepsilon$-approximate range-counting queries in $O((\log\Phi+\varepsilon^{-d})\lg^2\lg m)$ time. Theorem~2 (dynamic) uses the same asymptotic space. With $N=dbn$, insertions take $O(\log\Phi+b\log N/\log\log N+\log m)$ amortised time, deletions take $O((b\log N/\log\log N+\log m)\log\Phi)$ amortised time, and queries are slower by a factor $O(\log N/\log\log N)$.

\item \textbf{Usage of packing properties: Yes.}
In Section~5, before Lemma~4, they use the bound $C_{d,\varepsilon} = O(\varepsilon^{-d})$ on the number of leaves visited established by Arya et al.~\cite{Arya1998ANN}. 

\item \textbf{Result transfer: Yes for planar approximate nearest-neighbour functionality, with polylogarithmic overhead.} Theorem~\ref{thm:ann-recovery} gives $O(\log^4n+\varepsilon^{-2}\log n)$ queries and $O(\log^5n)$ amortised updates using $O(n\log^3n)$ space. The succinct bit-space representation does not transfer. 

\item \textbf{Compressed definition: No compressed links.} In Section~2 they specify an uncompressed PR-quadtree. In Section~6 they insert points by splitting leaves. Its space bound is linear in subdivision nodes, and not necessarily in the number of input points.
\end{itemize}

\subparagraph{Dynamic Data Structures for Approximate Hausdorff Distance in the Word-RAM~\cite{ChanSkrepetos2015}}
\begin{itemize}
\item \textbf{Model and assumptions.} In Section~1.1 they explicitly assume the Word-RAM on integer coordinates in $[U]$, with constant-time arithmetic, shifts, and logical operations on $\Theta(\log U)$-bit words. Grid cells are stored in hash tables (Section~2.1), which makes the update bound expected.

\item \textbf{Stated result.} For constant $d$ and $\varepsilon$, they maintain a $(1+\varepsilon)$-approximation of the Hausdorff distance between two dynamic point sets with $O(\log U/\log\log n)$ expected update time and constant query time.

\item \textbf{Usage of packing properties: Yes.} In Section~2.1 they observe that a cell of side length $r\varepsilon$ has a constant number of neighbouring cells within distance $r$ (depending only on $\varepsilon$ and $d$), which bounds how many counters an update changes. In Section~2.2.1 they organise the grids in an implicit quadtree.

\item \textbf{Result transfer: Yes for a weaker planar query bound.} We replace their grids by the WSPD of Theorem~\ref{thm:wspd} and use its pairs to recompute a $(1+\varepsilon)$-approximation of the directed Hausdorff distance from the blue point set to the red point set in $O(\varepsilon^{-2}n)$ time.
However, we do not establish a transfer of the dynamic update bound: in Section~2.2, they maintain per-level counters and packed emptiness flags across $O(\log U)$ grid levels. A static WSPD does not maintain those flags. 

\item \textbf{Compressed definition: Implicit canonical hierarchy.} In Section~2.2.1 they view the grids as an implicit, complete (uncompressed) quadtree of height $O(\log U)$. 
\end{itemize}

\subparagraph{Dynamic Streaming Algorithms for Geometric Independent Set~\cite{ChanYu2025}}
\begin{itemize}
\item \textbf{Model and assumptions.} The setting is a dynamic geometric stream: objects are inserted and deleted, and the guarantee concerns the working memory. In Section~1 and Theorem~9 they assume a bounded universe: the objects are contained in $[0,U)^d$, have diameter at least~$1$, and have weights in $[U]$. For rectangles (Theorem~5), the coordinates are integers in $[U]$.

\item \textbf{Stated result.} For any constant $\delta>0$, Theorem~9 gives an $O((1/\delta)^2)$-approximation of the maximum-weight independent set of fat objects in constant dimension, using $O(U^{O(\delta)}\log^{O(1)}n)$ space and update time with high probability. Theorem~5 gives the same bounds for the maximum independent set size of rectangles in the plane.

\item \textbf{Usage of packing properties: Yes.} In Lemma~8, they adapt the representative argument of Bhore and Chan~\cite{BhoreChan2025}. In Section~4 they assume comparable-size containment and fatness. The bound on the overlap among representatives is taken from Bhore and Chan.

\item \textbf{Result transfer: Not established.} The hashing and sampling tools of Sections~2--4 require finite encodings of objects. Replacing the geometric hierarchy by an explicit real-coordinate tree does not preserve the streaming space bound.

\item \textbf{Compressed definition: Canonical boxes in a bounded universe.} In Section~4 they use a hierarchy of quadtree boxes. In Footnote~5 they explain that the bounded universe allows a simpler divide-and-conquer than the balanced quadtree decomposition of Bhore and Chan.
\end{itemize}

\subsection{Explicit subdivision}
\label{app:literature-subdivision}
These algorithms explicitly subdivide cells. They either restrict the depth, pay for it through a dependence on spread, or analyse a local split rather than insertion of an arbitrary point. 
\subparagraph{Amortized Analysis of Smooth Quadtrees in All Dimensions~\cite{BennettYap2017}}
\begin{itemize}
\item \textbf{Model and assumptions.} The paper names no model of computation. The smooth split is a local operation (Section~1.1): the algorithm already holds a pointer to the leaf it splits. It uses only splits into $2^D$ congruent sub-boxes and pointers to principal neighbours.

\item \textbf{Stated result.} Theorem~1: any sequence of $n$ smooth splits, starting from a single box, performs at most $2^D(D+1)!\,n$ splits, so a smooth split takes amortised constant time for fixed~$D$. In Table~1, $n$ is the number of tree nodes, not of stored input points.

\item \textbf{Usage of packing properties: Yes, with smoothness.} In Section~1 they call a quadtree smooth if adjacent leaf boxes, that is, leaves sharing a facet, differ by at most one in height. In Section~1.1 they maintain this as an invariant. Smoothness bounds the number of leaves adjacent to a box, and after a smooth split only the neighbours of the split box can violate it, which makes the local repair of Algorithm~1 possible.

\item \textbf{Result transfer: Not established.} No point-insertion bound follows from Theorem~1: it charges per smooth split of a given leaf, and separating two arbitrarily close points may require arbitrarily many splits. The split itself needs only midpoint subdivision, pointer updates and depth comparisons, which our Real-RAM supports. Our tree does not maintain the smoothness invariant or the neighbour pointers that their theorem needs.

\item \textbf{Compressed definition: No compressed links.} In Section~1 an internal node is a box that is split into $2^D$ congruent sub-boxes. Compressed quadtrees appear only as related work (Section~1.4). The $\Theta(n)$ space in Table~1 is linear in the number of subdivision nodes. 
\end{itemize}

\subparagraph{Dynamic Approximate Maximum Independent Set of Intervals, Hypercubes and Hyperrectangles~\cite{Henzinger2020MIS}}
\begin{itemize}
\item \textbf{Model and assumptions.} The paper names no model of computation. All objects lie in a fixed $[0,N]^d$, where $N$ is given in advance, and have side lengths between $1$ and $N$. Weights, when present, lie in $[1,W]$. They state explicitly that coordinates need not be integers.

\item \textbf{Stated result.} Their algorithms are deterministic and fully dynamic (Table~1). For unweighted hypercubes they maintain a $(1+\varepsilon)2^d$-approximate independent set with worst-case update time $O_{d,\varepsilon}(\log^{2d+1}n\log^{2d+1}N)$. For weighted hypercubes, in Theorem~11 they give a $(4+O(\varepsilon))2^d$-approximation with worst-case update time in $O_{d,\varepsilon}(\log^{2d-1}n\log^{2d+1}N\log W)$.

\item \textbf{Usage of packing properties: Yes.} In Lemma~2 they bound the number of independent hypercubes assigned to one cell by $(d/\varepsilon)^d$: a hypercube is assigned to a cell only if its side length is at least an $\varepsilon/d$ fraction of the cell's. In the proof of Lemma~9 they use this bound to charge the light hypercubes in the approximation guarantee.

\item \textbf{Result transfer: Not established.} In Section~2 they assign each hypercube to a cell of a shifted grid over the fixed universe $[0,N]^d$, and in Lemma~1 they choose the shift. Their error analysis (Lemmas~7, 9 and~10) charges an $\varepsilon/\log N$ fraction per cell and sums over the $\log N$ levels. Our packing property replaces neither the assignment nor this analysis.

\item \textbf{Compressed definition: No compressed links.} In Section~2 the grid has $\log N$ levels, each halving the side length of the cells. An update recomputes the solutions of the at most $\log N$ cells that contain the object, so updates pay for every level.
\end{itemize}

\subparagraph{Dynamic Enumeration of Similarity Joins~\cite{Agarwal2021Similarity}}
\begin{itemize}
\item \textbf{Model and assumptions.} The paper names no model of computation. In Section~3.1 they assume that the spread of $A\cup B$, the ratio of the largest to the smallest interpoint distance, is $n^{O(1)}$.

\item \textbf{Stated result.} Theorem~8: for constant $d$ and polynomially bounded spread, a structure of $O(\varepsilon^{-d}n)$ space is updated in $\widetilde O(\varepsilon^{-d})$ time and supports $\varepsilon$-approximate enumeration of the similarity join with $O(1)$ delay, for any query threshold $r$ and any $\ell_p$ metric. That is, it reports every pair at distance at most $r$, and possibly pairs at distance up to $(1+\varepsilon)r$. Here $\widetilde O$ hides polylogarithmic factors in $n$. 
They state no bound for larger spread.

\item \textbf{Usage of packing properties: Yes.} In Section~3.1 they quote that, under polynomial spread, the quadtree yields a WSPD with $O(\varepsilon^{-d}n)$ pairs in which each cell appears in $O(\varepsilon^{-d}\log n)$ pairs. These bounds control the number of stored pairs and the number of pairs that change in an update.

\item \textbf{Result transfer: Not established.} Theorem~\ref{thm:wspd} supplies a static planar WSPD. For updates they rely on the dynamic WSPD techniques of Callahan~\cite{Callahan1995Thesis} and of Fischer and Har-Peled~\cite{FischerHarPeled2005}, which we discuss in Appendix~\ref{app:callahan} and Appendix~\ref{app:literature-gaps}. They must also maintain the ordered set of pairs and the pointers and leaf lists used for enumeration. 

\item \textbf{Compressed definition: No compressed links.} In Section~3.1 the quadtree subdivides each cell into $2^d$ congruent cells until a cell contains at most one point, and they invoke polynomial spread for $O(\log n)$ height. No compressed quadtree is defined.
\end{itemize}

\subparagraph{Dynamic Mesh Refinement with Quad Trees and Off-Centers~\cite{AcarHudson2008}}
\begin{itemize}
\item \textbf{Model and assumptions.} The paper names no model of computation. The points lie in a bounding box $[0,L]^d$. The dynamic algorithm is obtained from the self-adjusting-computation framework (Section~4) and is not written out.

\item \textbf{Stated result.} Theorem~4.6: a well-graded quadtree is maintained under point insertions and deletions in $O(\log(L/s))$ time per update and $O(n\log(L/s))$ space, where $s$ is the closest-pair distance, so that $L/s$ is the spread.

\item \textbf{Usage of packing properties: Yes.} In Lemma~2.3 they bound the neighbours of a cell by a volume argument. In Lemmas~3.1 and~3.2 they show that an update causes $O(1)$ splits per size class, all near the changed point, hence $O(\log(L/s))$ in total.

\item \textbf{Result transfer: Not established.} Their quadtree operations already accept real-valued input, at a cost of $O(\log(L/s))$ per update. Our packing property does not remove this dependence on the spread.

\item \textbf{Compressed definition: No compressed links.} In Sections~2--4 they refine a well-graded quadtree by explicit splits, and Theorem~4.6 pays for the depth in both time and space.
\end{itemize}

\subparagraph{Dynamic Well-Spaced Point Sets~\cite{Acar2013WellSpaced}}
\begin{itemize}
\item \textbf{Model and assumptions.} The paper names no model of computation. Beyond quadtree splits and pointer traversal, the superset algorithm computes clipped Voronoi cells (Section~5) and ranks vertices by $\lfloor\log_\rho\rfloor$ of their nearest-neighbour distance (Section~6).

\item \textbf{Stated result.} They maintain a size-optimal well-spaced superset of a dynamic point set in $O(\log\Delta)$ time per insertion or deletion, where $\Delta$ is the geometric spread. Theorem~4.3 gives the same bound for updating the underlying quadtree.

\item \textbf{Usage of packing properties: Yes.} In Lemmas~4.1 and~4.2 they show that every square changed by an update lies within distance proportional to its side length of the updated point. With the crowding and grading conditions this gives $O(1)$ operations per level, hence Theorem~4.3.

\item \textbf{Result transfer: Not established.} Their quadtree operations already accept real-valued input, at a cost of $O(\log\Delta)$ per update. By Theorem~11.2, $n$ insertions can force $\Omega(n\log\Delta)$ new Steiner vertices, so no algorithm that maintains the superset explicitly is independent of the spread.

\item \textbf{Compressed definition: No compressed links.} In Section~4 they descend and split cells explicitly, paying for every level.
\end{itemize}

\subsection{Restricted constructions}
\label{app:literature-restrictions}
These constructions fix a range of object sizes in advance, or restrict compressed links.
\subparagraph{Dynamic Planar {Voronoi} Diagrams for General Distance Functions and Their Algorithmic Applications~\cite{Kaplan2020Voronoi}}
\begin{itemize}
\item \textbf{Model and assumptions.} They work in an algebraic model of computation for the distance functions (Section~3). In Section~9.2 they assume that all disk radii lie in $[1,\Psi]$.

\item \textbf{Stated result.} Theorem~9.11: connectivity among disks is maintained in $O(\Psi^2\log^9 n\,\lambda_6(\log n))$ expected amortised update time and $O(\log n/\log\log n)$ query time, where $\lambda_6$ is the maximum length of a Davenport--Schinzel sequence of order six.

\item \textbf{Usage of packing properties: Yes, for a grid.} In Section~9.2 they place the disk centres in a uniform grid of diameter-one cells, in which the neighbourhood of a cell consists of $O(\Psi^2)$ cells. In Lemma~9.10 they maintain a maximal matching between the disks of two neighbouring cells.

\item \textbf{Result transfer: Possibly.} We believe that our new quadtree may support the required packing properties at a slowdown, as we establish the key packing property that is used in this paper. To actually transfer the result, we would need to implement all their auxiliary data structures without the assumption that the disks have fixed and bounded radii. For now, we consider this task out of scope. 

\item \textbf{Compressed definition: Uniform grid.} In Section~9.2 they use a single grid of diameter-one cells. The lookup of the cell containing a point is left open.
\end{itemize}

\subparagraph{Dynamic Disk Connectivity~\cite{KauerMulzer2019Disks}}
\begin{itemize}
\item \textbf{Model and assumptions.} The extended abstract names no model of computation. Radii lie in $[\phi,1]$ for a fixed $0<\phi\leq1$.

\item \textbf{Stated result.} Theorem~3.7: $O(\phi^{-1}2^{\alpha(n)}\log^{10}n)$ expected amortised update time, $O(\log n)$ amortised query time and $O(\phi^{-1}n\log^3 n)$ expected space, where $\alpha$ is the inverse Ackermann function.

\item \textbf{Usage of packing properties: Yes.} In Equation~(1) they count $O(\phi^{-2})$ neighbouring cells, and in Lemma~3.2 they bound the cells near the boundary of a disk of radius $r$ by $O(r/\phi)$.

\item \textbf{Result transfer: Possibly.} We believe that our new quadtree may support the required packing properties at a slowdown, as we establish the key packing property that is used in this paper. To actually transfer the result, we would need to implement all their auxiliary data structures without the assumption that the disks have fixed and bounded radii. For now, we consider this task out of scope. 

\item \textbf{Compressed definition: No compressed links.} In Section~2 they use quadtrees of height at most $\lceil\log(1/\phi)\rceil$, rooted at the cells of a uniform grid, and create nodes on first access.
\end{itemize}

\subparagraph{Dynamic Connectivity in Disk Graphs~\cite{Baumann2024Disks}}
\begin{itemize}
\item \textbf{Model and assumptions.} In Section~2.5 they work on a Real-RAM with a floor function, and in Appendix~A they remove the floor by aligning quadtrees locally. For the fully dynamic result, radii lie in $[1,\Psi]$ with $\Psi$ known in advance.

\item \textbf{Stated result.} Theorem~4.6: $O(\Psi\log^4 n)$ expected amortised update time, $O(\log n/\log\log n)$ query time and $O(\Psi n\log n)$ space.

\item \textbf{Usage of packing properties: Yes.} In Lemma~4.1 they bound the nearby cells of a cell by $O(\Psi^2)$. In Lemma~4.4 they count the cells along the boundary of a disk, which gives the linear dependence on~$\Psi$.

\item \textbf{Result transfer: Possibly.} We believe that our new quadtree may support the required packing properties at a slowdown, as we establish the key packing property that is used in this paper. To actually transfer the result, we would need to implement all their auxiliary data structures without the assumption that the disks have fixed and bounded radii. For now, we consider this task out of scope. 

\item \textbf{Compressed definition: Local alignment.} In Appendix~A.1 they build quadforests of bounded height inside squares whose size depends on~$\Psi$. In Appendix~A.2 they build a compressed quadtree once, for the decremental result with arbitrary radii.
\end{itemize}

\subparagraph{A Dynamic $(1+\varepsilon)$-Spanner for Disk Intersection Graphs~\cite{deBerg2026DiskSpannerFull}}
\begin{itemize}
\item \textbf{Model and assumptions.} They state that their results apply to the Word-RAM, the Real-RAM and the pointer machine. All diameters lie in $[4,\Psi]$ for a fixed, known~$\Psi$.

\item \textbf{Stated result.} Theorem~1: a $(1+\varepsilon)$-spanner of the disk intersection graph is maintained in $O((\Psi/\varepsilon)^2\log^4 n\log^2\Psi\log^2(\varepsilon^{-1}))$ expected amortised update time and $O(n\varepsilon^{-2}\log^4 n\log\Psi)$ expected space. Theorem~2: connectivity is maintained in $O(\Psi\log^4 n\log\Psi)$ expected amortised update time and $O(n\log^4 n\log\Psi)$ expected space.

\item \textbf{Usage of packing properties: Yes.} In the proof of Lemma~20 they count, for the cell storing a disk, the $O(\log\Psi)$ nearby larger cells, which bounds the number of matching edges. The spanner also uses the storing families of Definition~3 and Euclidean spanners between nearby cells.

\item \textbf{Result transfer: Possibly for connectivity.} We believe that our new quadtree may support the required packing properties at a slowdown, as we establish the key packing property that is used in this paper. To actually transfer the result, we would need to implement all their auxiliary data structures without the assumption that the disks have fixed and bounded radii. For now, we consider this task out of scope. 

\item \textbf{Compressed definition: Local decompositions of fixed size.} In Appendix~B they describe compression by minimum bounding hypercubes for the static case. For updates they use the focused spanner decompositions of Appendices~B.1 and~B.2, which keep uncompressed quadtrees on boxes of size proportional to~$\Psi$.
\end{itemize}

\subparagraph{Dynamic Smooth Compressed Quadtrees~\cite{Hoog2018Smooth} \\ On Modelling, Structuring and Capturing Geometric Information~\cite{Hoog2022Thesis}}
\begin{itemize}
\item \textbf{Model and assumptions.} The SoCG paper claims its result for arbitrary point sets on a Real-RAM pointer machine~\cite{Hoog2018Smooth}. The thesis states the result under the assumption that the root of a new uncompressed component is computed in constant time, and notes that real-valued point sets can violate this~\cite[Theorem~7.7 and Section~7.7]{Hoog2022Thesis}. It gives two restrictions under which the assumption holds: constantly many compressed links, or at most one compressed link on every root-to-leaf path.

\item \textbf{Stated result.} For constant dimension, a smooth compressed quadtree is maintained under insertions and deletions of points in worst-case constant time per update, given a pointer to the cell of the update~\cite{Hoog2018Smooth}.

\item \textbf{Usage of packing properties: Yes, with smoothness.} In Definition~4 of the SoCG paper a cell is smooth if its neighbouring leaves are at most a constant factor larger, and in Observation~1 they count the equal-size neighbours of a cell. Smoothness bounds the neighbours that an update processes.

\item \textbf{Result transfer: Not established.} Our tree does not maintain smoothness or neighbour pointers.

\item \textbf{Compressed definition: Non-aligned components.} An uncompressed component may start at any scale. The thesis realigns a component with its parent gradually, under the two restrictions above.
\end{itemize}

\subsection{Additional real-coordinate operations}
\label{app:literature-strong}
The papers in this subsection work with real coordinates but allow more
than the Real-RAM does: a floor function, bitwise operations on the binary
expansions of reals, or a primitive that returns canonical quadtree cells.
Such operations are far from innocent. Sch\"onhage~\cite{Schonhage1979RAM}
showed that a unit-cost RAM with addition, multiplication and integer
division decides every problem in NP in polynomial time. With bitwise
Boolean operations in place of division, such a machine decides exactly
the problems in
PSPACE~\cite{HartmanisSimon1974,PrattRabinStockmeyer1974}. Both make such machines unreasonably strong.
Operations on words of logarithmic length do not have this effect.

For each paper we record which operations it assumes. Some assume floor
outright. For the others we show below that floor can be recovered from
what they assume.
The bullets labelled ``Power of the model'' say which of the two results
applies: NP when the operations of a paper yield floor, and PSPACE when
they yield bitwise-and.
When a paper offers both a Word-RAM and a Real-RAM version, we discuss
the latter.

As defined in the introduction, a compressed link from a cell $C$ to a cell $C'$ is canonical if $C'$ can be obtained from $C$ by a sequence of regular splits. We call a compressed quadtree canonical if all its compressed links are canonical. Additionally, we call its root canonical if it belongs to the standard dyadic hierarchy, consisting of the half-open squares
\[
[i2^j,(i+1)2^j)\times[k2^j,(k+1)2^j),
\qquad i,k,j\in\mathbb Z.
\]
Thus, in a canonical compressed quadtree with a canonical root, every cell belongs to this hierarchy. As before, we assume that the quadtree explicitly stores the coordinates of each cell.

\begin{theorem}[Canonical quadtrees implement floor]
\label{thm:canonical-floor}
Consider a data structure that maintains a canonical compressed quadtree
of linear size under point insertions in $O(\log(n+2))$ time.
If it maintains a canonical root for arbitrary positive real-valued point coordinates, regardless of their magnitude or precision, then it can be used to compute $\lfloor x\rfloor$ for any real $x$ in $O(1)$ time.
If instead its root is fixed to $[0,1)^2$, it can be used to compute
$\lfloor x\rfloor$ in $O(1)$ time for any $x\geq0$, given a power of two
$M>x+2$.
\end{theorem}

\begin{proof}
Let $z\geq1$. We insert the three points
$(z-1/2,1/4)$, $(z,1/4)$, and $(z+1/2,1/4)$ into an empty tree. Consider first the case that the quadtree has a canonical root (such a root exists as all coordinates are positive).
The unit square $S=[\lfloor z\rfloor,\lfloor z\rfloor+1)\times[0,1)$ contains $(z,1/4)$.
If $z<\lfloor z\rfloor+1/2$, this point $(z,1/4)$ lies in the left half of $S$ and
$(z+1/2,1/4)$ lies in the right half. Otherwise $(z,1/4)$ lies in the
right half and $(z-1/2,1/4)$ in the left half.
Either way two children of $S$ are occupied, and thus the cell $S$ must be in the quadtree.
The tree has $O(1)$ nodes. We look through them for the cell of side
length one that contains $(z,1/4)$ in constant time. 

For a real $x$, we apply this to $z=|x|+1$ and subtract one, which gives
$\lfloor|x|\rfloor$ from which we obtain $\lfloor x \rfloor$.

Now, consider the case that the root is fixed to $[0,1)^2$. We take $z=x+1$ and divide all
coordinates by $M$. The three points lie in $[0,1)^2$ because $M>x+2$.
Since $M$ is a power of two, the scaled copy of $S$ is again a dyadic
square, now of side length $1/M$, and it is a branching cell for the
same reason. We find it among the $O(1)$ nodes, multiply its left
boundary by $M$ and subtract one.
\end{proof}

\subparagraph*{From bit operations to floor.}
\label{app:floor-audit}
Several papers do not assume access to the floor operation. However, they \emph{do} allow bitwise operations on reals.
Let $M$ be a power of two and $0\leq x<M$. 
We write numbers in binary, choosing the expansion ending in zeros whenever there is a choice. For example, we write $1/2$ as $0.1000\ldots$ rather than $0.0111\ldots$.
Then
\[
\lfloor x\rfloor
= M\bigl((x/M)\mathbin{\mathrm{AND}}(1-1/M)\bigr).
\]
In binary, $1-1/M$ has ones in the first $k$ positions after the binary point and zeros thereafter. Thus the AND keeps the first $k$ bits of $x/M$, and multiplying by $M$ gives the integer part of $x$.
The mask $1-1/M$ keeps exactly the leading $\log_2M$ bits of $x/M$, and
these are the bits of $\lfloor x\rfloor$. Both operands lie in $[0,1)$.
In the same way, the bitwise-and of two integers $u,v<M$ is
$M\bigl((u/M)\mathbin{\mathrm{AND}}(v/M)\bigr)$, so bitwise-and on reals
in $[0,1)$ is as strong as on integers.
Exclusive-or is enough as well, since
$u\mathbin{\mathrm{AND}}v=(u+v-(u\ \mathbin{\mathrm{XOR}}\  v))/2$.
An operation that only reports the position of the first differing bit
is not enough for this argument: we need the masked value, or the cell
boundary, as a number.

\subparagraph*{Bounded floor.}
The formula above and the second part of
Theorem~\ref{thm:canonical-floor} compute $\lfloor x\rfloor$ only when a
power of two larger than $x$ is known. This restriction costs little.
Consider a computation with $t$ arithmetic and floor operations on
rational inputs of total bit length $L$. An arithmetic operation at most
doubles the bit length of the numbers involved, up to an additive
constant, and floor does not increase it. All intermediate values are
therefore smaller than $2^{2^{O(t+L)}}$, and squaring the number $2$
repeatedly, $O(t+L)$ times, gives a power of two of that size. With this
$M$ we can simulate every floor in the computation, at polynomial
overhead. Negative arguments are handled as in
Theorem~\ref{thm:canonical-floor}.

\subparagraph{Near-Optimal Dynamic Steiner Spanners for Constant-Curvature Spaces~\cite{KisfaludiBak2026Steiner}}
\begin{itemize}
\item \textbf{Model and assumptions.} In Section~2 they state that their results require floor and bitwise operations, and that they otherwise hold on both the Real-RAM and the Word-RAM.

\item \textbf{Stated result.} Theorem~1: for $n$ points in $d$-dimensional Euclidean, spherical or hyperbolic space, they give a Steiner $(1+\varepsilon)$-spanner with $O_d(\varepsilon^{(1-d)/2}\log(1/\varepsilon)\,n)$ edges, maintained in $O_d(\varepsilon^{(1-d)/2}\log(1/\varepsilon)\log n)$ time per insertion or deletion.

\item \textbf{Power of the model: At least NP.} Floor gives NP. PSPACE follows if their bitwise operations include bitwise-and or exclusive-or.

\item \textbf{Usage of packing properties: Shifted-cell containment.} In Section~3 they use property~$(*)$: every pair of points lies in a cell of one of $O(d)$ shifted quadtrees whose diameter is within a factor $O(d\sqrt d)$ of their distance. The construction preceding Theorem~5 also places Steiner points on cell boundaries.

\item \textbf{Result transfer: Not established.} Our packing property does not supply property~$(*)$, the boundary construction, or the spherical and hyperbolic hierarchies.

\item \textbf{Compressed definition: Implicit canonical hierarchy.} The dynamic structure keeps the points in sorted lists, one per shifted quadtree, and compares points directly in these orders.
\end{itemize}

\subparagraph{Dynamic Geometric Independent Set~\cite{Bhore2020MIS}}
\begin{itemize}
\item \textbf{Model and assumptions.} The paper names no model of computation for squares. In Footnote~3 they note that finding the smallest quadtree node containing a square can be implemented with a binary logarithm, a floor and an exponentiation.

\item \textbf{Stated result.} Theorem~1.2: they maintain an expected $O(1)$-approximate maximum independent set of axis-aligned squares in $O(\log^5n)$ amortised time per update. For hypercubes they give an $O(4^d)$-approximation in $O(2^d\log^{2d+1}n)$ time.

\item \textbf{Power of the model: NP.} They use floor.

\item \textbf{Usage of packing properties: Shifted containment and path structure.} In Lemma~4.1 they show that, in a randomly shifted and scaled quadtree, the squares that contain the centre of their assigned cell retain a constant fraction of an optimum in expectation. In Lemma~4.2 they combine independent sets along paths of the quadtree.

\item \textbf{Result transfer: Not established.} Our tree does not preserve the random cell assignment of Lemma~4.1 or the path decomposition of Lemma~4.2.

\item \textbf{Compressed definition: Compressed canonical hierarchy.} In Section~5.3 they maintain a compressed version of the randomly shifted quadtree. All its cells are cells of that hierarchy.
\end{itemize}

\subparagraph{Dynamic Compressed Hyperoctrees with Application to the {N}-body Problem~\cite{AluruSevilgen1999}}
\begin{itemize}
\item \textbf{Model and assumptions.} In Section~1 they assume an extended model in which floor, bitwise exclusive-or and logarithm take constant time. They use it in Lemma~2, which they attribute to Clarkson, to find the smallest cell containing a box.

\item \textbf{Stated result.} Theorem~1: they build a compressed hyperoctree of $n$ points in $O(n\log n)$ time. Theorem~2: point search, cell search, insertion and deletion take $O(\log n)$ time.

\item \textbf{Power of the model: PSPACE.} On real coordinates, exclusive-or yields bitwise-and.

\item \textbf{Usage of packing properties: No.} The linear size follows because every internal node has at least two children (Section~2).

\item \textbf{Result transfer: Yes for a linear-size planar point tree with logarithmic updates and point location,} by Theorem~\ref{thm:dynamic} and Appendix~\ref{app:edge-oracles}. Their cell search and their $N$-body result remain specific to their tree.

\item \textbf{Compressed definition: Canonical cells.} In Section~2 each node stores a large cell $L(v)$ and the smallest cell $S(v)$ of the hierarchy that contains the same points. Children arise by splitting $S(v)$.
\end{itemize}

\subparagraph{Skip Quadtrees: Dynamic Data Structures for Multidimensional Point Sets~\cite{Eppstein2008Skip}}
\begin{itemize}
\item \textbf{Model and assumptions.} In Section~2.2 they assume constant-time operations on point coordinates, such as finding the most significant binary digit in which two coordinates differ. For other number representations they mention floor and ceiling. Points are in $\mathbb R^d$, and each cell is stored by its centre and half side length.

\item \textbf{Stated result.} Theorems~2 and~3: search, insertion and deletion take $O(\log n)$ time, with high probability for the randomized and in the worst case for the deterministic skip quadtree, using $O(n)$ space. Theorem~4: approximate range reporting in $O(\log n+\varepsilon^{1-d}+k)$ time for output size~$k$. Theorem~5: approximate nearest-neighbour queries in $O(\varepsilon^{1-d}(\log n+\log(1/\varepsilon)))$ time.

\item \textbf{Power of the model: NP.} The cells are stored with numerical coordinates, so Theorem~\ref{thm:canonical-floor} yields floor.

\item \textbf{Usage of packing properties: Yes.} In Section~5.3 (Lemmas~4 and~5) they bound the number of cells of a given size that a range query visits, using the volume of the thin region around the boundary of the query range.

\item \textbf{Result transfer: Yes for planar approximate nearest neighbours, with polylogarithmic overhead for fixed~$\varepsilon$.} Theorem~\ref{thm:ann-recovery}  answers approximate nearest neighbour queries in $O(\log^4n+\varepsilon^{-2}\log n)$, with $O(\log^5n)$ amortised updates and $O(n\log^3n)$ space. Their linear space, their update bound and their dependence on~$\varepsilon$ are stronger. 

\item \textbf{Compressed definition: Canonical cells.} In Section~2.3 the compressed quadtree keeps the ``interesting'' squares of the standard quadtree over a given root square.
\end{itemize}

\subparagraph{A Dynamic Data Structure for Approximate Range Searching~\cite{MountPark2010}\\ A Self-adjusting Data Structure for Multidimensional Point Sets~\cite{ParkMount2012}\\ Dynamic Data Structures for Geometric Search and Retrieval~\cite{Park2013Thesis}}
\begin{itemize}
\item \textbf{Model and assumptions.} Mount and Park assume a model of computation that supports exclusive-or and base-2 logarithms on point coordinates, so that the smallest quadtree box enclosing two points is found in constant time~\cite[Section~2]{MountPark2010}. The thesis repeats this~\cite[p.~56]{Park2013Thesis}. Points are scaled into the unit hypercube. Park and Mount~\cite{ParkMount2012} take their insertion from the earlier paper.

\item \textbf{Stated result.} The quadtreap~\cite{MountPark2010} uses $O(n)$ space and has height $h=O(\log n)$ with high probability. Insertions take $O(h)$ time, deletions $O(h)$ expected time, and approximate range-counting and nearest-neighbour queries $O(h+\varepsilon^{1-d})$ time. The splay quadtree~\cite{ParkMount2012} supports insertions and deletions in $O(\log n)$ amortised time.

\item \textbf{Power of the model: PSPACE.} Exclusive-or yields bitwise-and.

\item \textbf{Usage of packing properties: Yes.} In Theorem~3.4.1 of the thesis (pp.~93--97), the query analysis counts the large cells meeting the range and bounds the smaller cells by size class. The conference paper takes this count from Arya and Mount~\cite{AryaMount2000Range}.

\item \textbf{Result transfer: Not Established} 
Their rotations and shrinking operations, their general convex ranges and their working-set bounds are specific to their trees.

\item \textbf{Compressed definition: Canonical cells.} They use the quadtree boxes of a binary subdivision of the unit hypercube, together with shrinking. The smallest enclosing box is a box of this hierarchy.
\end{itemize}

\subparagraph{Approximating the $\lambda$-low-density value~\cite{GudmundssonHuangWong2023}}
\begin{itemize}
\item \textbf{Model and assumptions.} They refer to ``the unit RAM model'' (Lemma~8), in which they find the first bit where two coordinates in $[0,1)$ differ and truncate there. For insertions (Appendix~B) they rely on Har-Peled's Corollary~2.17~\cite{HarPeled2011}, which they quote with the hypothesis that floor and the first-differing-bit operation take constant time.

\item \textbf{Stated result.} They compute a $3$-approximation of the $\lambda$-low-density value of $n$ segments in the plane in $O(n\log n+\lambda n)$ time, and maintain it under insertions in $O(\log n+\lambda^2)$ amortised time per segment (Theorem~15).

\item \textbf{Power of the model: NP.} Their insertion algorithm depends on floor.

\item \textbf{Usage of packing properties: Yes, together with a covering.} In Lemma~7 they cover the neighbourhood of a segment by constantly many canonical squares of comparable size. In Corollary~11 they use low density to bound the longer segments meeting such a neighbourhood by $O(\lambda)$.

\item \textbf{Result transfer: Not established.} Our packing property does not construct the canonical covers of Lemma~7 or maintain the per-node segment lists of their insertion algorithm.

\item \textbf{Compressed definition: Canonical cells.} Their squares have dyadic corners, and the tree is Har-Peled's compressed quadtree over these squares.
\end{itemize}

\subparagraph{On Locality-Sensitive Orderings and Their Applications~\cite{chan2019Locality}}
\begin{itemize}
\item \textbf{Model and assumptions.} They assume a unit-cost Real-RAM with arithmetic and comparisons, explicitly without floor, augmented with bitwise exclusive-or and bitwise-and (p.~21:5). In Lemma~9 they apply these to real coordinates to compare two points in an ordering.

\item \textbf{Stated result.} Theorem~10: there are $O(\varepsilon^{-d}\log(1/\varepsilon))$ orderings of $[0,1)^d$ such that, for any two points $p,q$, in one of them every point between $p$ and $q$ is within distance $\varepsilon\|p-q\|$ of $p$ or of $q$. Theorems~12 and~14: they maintain a $(1+\varepsilon)$-approximate bichromatic closest pair and a $(1+\varepsilon)$-spanner with $O(n\varepsilon^{-d}\log(1/\varepsilon))$ edges in $O(\varepsilon^{-d}\log^2(1/\varepsilon)\log n)$ time per update.

\item \textbf{Power of the model: PSPACE.} They assume bitwise-and.

\item \textbf{Usage of packing properties: Shifted-cell containment.} In Lemma~7 they show that every pair of points lies in a cell of comparable size in one of $2\lceil d/2\rceil+1$ shifted quadtrees.

\item \textbf{Result transfer: Not established.} Our packing property does not supply the shifted containment of Lemma~7 or their orderings, which are fixed in advance.

\item \textbf{Compressed definition: Implicit canonical hierarchy.} The points are kept in search trees sorted by the orderings, and the quadtree is implicit in these orderings.
\end{itemize}

\subparagraph{Near-Optimal Euclidean Locality-Sensitive Orderings~\cite{GaoHarPeled2025LSO}}
\begin{itemize}
\item \textbf{Model and assumptions.} They assume a unit-cost Real-RAM with arithmetic, comparisons and the floor function, augmented with bitwise exclusive-or and bitwise-and (p.~3).

\item \textbf{Stated result.} Theorem~3.7: they reduce the number of orderings to $O(\varepsilon^{1-d}\log(1/\varepsilon))$. Theorems~4.3 and~4.4: they maintain a $(1+\varepsilon)$-approximate bichromatic closest pair and a $(1+\varepsilon)$-spanner with $O(n\varepsilon^{1-d}\log(1/\varepsilon))$ edges in $O(\varepsilon^{1-d}\log^2(1/\varepsilon)\log n)$ time per update.

\item \textbf{Power of the model: PSPACE.} They assume bitwise-and and floor.

\item \textbf{Usage of packing properties: Yes, together with sphere packings.} In Lemma~2.10 they construct a packing of the unit sphere, and in Lemma~3.3 they partition a packing into more widely separated classes. The proof of Theorem~3.7 also uses shifted-cell containment.

\item \textbf{Result transfer: Not established.} Our packing property does not supply the shifted containment or orderings that are fixed in advance.

\item \textbf{Compressed definition: Implicit canonical hierarchy.} The orderings are defined on quadtrees with fixed roots.
\end{itemize}

\subsection{Unspecified operations}
\label{app:literature-gaps}
These papers prescribe a quadtree construction and leave its implementation on the Real-RAM open, or take the construction from another paper. For the online results, the competitive guarantee is separate from the running time.
\subparagraph{Dynamic Stabbing Queries with Sub-logarithmic Local Updates for Overlapping Intervals~\cite{KhramtcovaLoffler2017}}
\begin{itemize}
\item \textbf{Model and assumptions.} The paper names no model of computation. The intervals are real and lie in a bounding interval that can be enlarged.

\item \textbf{Stated result.} Theorem~3: with $O(n)$ space, stabbing queries, insertions and deletions take $O(\log n+k\log n/\log\log n)$ time and local updates take $O(k\log n/\log\log n)$ time, where $k$ is the current ply. The insertion bound is amortised.

\item \textbf{Usage of packing properties: Yes, an overlap bound.} In Section~3 they use the ply to bound the intervals that can contain a query point, so that $O(k)$ marked-ancestor queries suffice.

\item \textbf{Result transfer: Not established.} Our packing property does not supply their interval storage or their marked-ancestor structures.

\item \textbf{Compressed definition: Taken from L\"offler, Simons and Strash.} In Section~4 they use $a$-compressed cells and assume decompression in constant time, citing~\cite{LofflerSimonsStrashFull}.
\end{itemize}

\subparagraph{Approximate Stabbing Queries with Sub-logarithmic Local Replacement~\cite{HoogLoffler2018Stabbing}}
\begin{itemize}
\item \textbf{Model and assumptions.} The extended abstract names no model of computation. In Section~2 they assume a dynamic compressed quadtree with $O(dn)$ space, logarithmic point location and constant-time navigation between cells of similar size, and refer to a full version for it.

\item \textbf{Stated result.} Theorem~5: for spheres or axis-parallel hypercubes of ply $k$ in $\mathbb R^d$, with $O(dn)$ space, insertions, deletions and $2^m$-approximate stabbing queries take $O(3^dk\log n/\log\log n+\log n+m)$ time, and local replacements take $O(2^{(d-1)m}k\log n/\log\log n)$ time. The proofs for $d>1$ are deferred to the full version.

\item \textbf{Usage of packing properties: Yes, an overlap bound.} In Sections~3 and~4 they bound the candidate regions in each direction by the ply.

\item \textbf{Result transfer: Not established.} Our packing property does not supply the assumed navigation or the marked-ancestor structures.

\item \textbf{Compressed definition: Assumed.} In Section~2 a region is stored in the largest cell that it contains and that contains its centre. The construction of compressed cells is left to the full version.
\end{itemize}

\subparagraph{Dynamic Well-Separated Pair Decomposition Made Easy~\cite{FischerHarPeled2005}}
\begin{itemize}
\item \textbf{Model and assumptions.} The paper names no model of computation. The points lie in $[0,1]^d$. In an insertion, the cell of a new branching node is created in constant time once its position in the tree is known (end of Section~2); the computation of this cell is left open.

\item \textbf{Stated result.} Theorem~5: an $\varepsilon^{-1}$-WSPD of size $O(n\varepsilon^{-d})$ is constructed in $O(n\log n+n\varepsilon^{-d})$ time. Theorems~12 and~13: with high probability, the list of pairs is updated in $O((\log n+\log\varepsilon^{-1})\varepsilon^{-d}\log n)$ time per insertion or deletion.

\item \textbf{Usage of packing properties: Yes, through Callahan and Kosaraju.} For the size bound of Theorem~5 they refer to the argument of Callahan and Kosaraju. In Sections~4 and~5 they shift the quadtree randomly to bound the levels searched during an update.

\item \textbf{Result transfer: Not established for updates.} Theorem~\ref{thm:wspd} constructs a static planar WSPD on our tree.

\item \textbf{Compressed definition: Canonical cells.} In Definitions~1 and~2 the compressed quadtree is obtained from a quadtree by removing empty leaves and contracting paths of single children. In Section~3.2 they navigate it with topology trees.
\end{itemize}

\subparagraph{Fully-Adaptive Dynamic Connectivity of Square Intersection Graphs~\cite{Hoog2024SquareConnectivity}}
\begin{itemize}
\item \textbf{Model and assumptions.} The paper names no model of computation. The aspect ratio $\psi$ is the ratio of the largest to the smallest square diameter at the time of the update. For insertions, deletions and point location in the compressed quadtree they refer to lecture notes of Har-Peled~\cite{HarPeled2010Quadtrees} (Section~4).

\item \textbf{Stated result.} Theorem~10 with Lemma~5: connectivity among axis-aligned squares is maintained in $O(\psi\log^4 n+\log^6 n)$ amortised update time, $O(\log n/\log\log n)$ query time and $O(n\log^3 n\log\psi)$ space.

\item \textbf{Usage of packing properties: Yes.} In Lemma~3 they count the cells that can store a square covering an ancestor of a given cell: $O(1)$ per level over $O(\log\psi)$ levels. In Lemma~5 they bound the cells along the boundary of a square by $O(\psi)$.

\item \textbf{Result transfer: Possibly.} We believe that our new quadtree may support the required packing properties at a slowdown, as we establish the key packing property that is used in this paper. To actually transfer the result, we would need to implement all their auxiliary data structures without the assumption that the disks have fixed and bounded radii. For now, we consider this task out of scope. 

\item \textbf{Compressed definition: Canonical cells.} In Section~3 they define the cells from a bounding square whose coordinates are powers of two, and use $\alpha$-compressed quadtrees.
\end{itemize}

\subparagraph{Fast Static and Dynamic Approximation Algorithms for Geometric Optimization Problems: Piercing, Independent Set, Vertex Cover, and Matching~\cite{BhoreChan2025}}
\begin{itemize}
\item \textbf{Model and assumptions.} The paper names no model of computation. For fat objects (Sections~2.2 and~3.2) they use shifted quadtree boxes. For vertex cover and matching (Sections~4--6) they use a dynamic structure for intersection detection.

\item \textbf{Stated result.} Theorems~2.11 and~3.6: for fat objects in constant dimension, an $O(1/\delta)$-approximation of the minimum piercing set and of the maximum-weight independent set is maintained in $O(n^{O(\delta)})$ amortised update time. Corollaries~4.5, 5.3 and~6.4: $(1+\varepsilon)$-approximate vertex cover and matching for disks are maintained in polylogarithmic amortised update time.

\item \textbf{Usage of packing properties: Yes, with containment.} In Definition~2.8 an object is good if it lies in a quadtree box of comparable size. In the proof of Lemma~2.10 they use this to show that the objects crossing a cell boundary are large, and fatness then bounds the piercing points.

\item \textbf{Result transfer: Not established for piercing and independent set.} Our packing property does not supply the containment of Definition~2.8. Their vertex cover and matching results for disks use an intersection oracle and need no quadtree.

\item \textbf{Compressed definition: Canonical cells.} In Definition~2.6 they define quadtree boxes, and in Fact~2.7 they partition space into quadtree boxes and differences of two such boxes, as in a compressed quadtree.
\end{itemize}

\subparagraph{Online Euclidean Spanners~\cite{BhoreToth2021Online}}
\begin{itemize}
\item \textbf{Model and assumptions.} The paper names no model of computation. Its guarantee is a competitive ratio. For the running time they use the incremental algorithm of Fischer and Har-Peled~\cite{FischerHarPeled2005} and quote its expected update time (Section~3).

\item \textbf{Stated result.} Theorem~4: the online algorithm has competitive ratio $O(\varepsilon^{-(d+1)}\log n)$ for $n$ points in $\mathbb R^d$.

\item \textbf{Usage of packing properties: Yes.} In the proof of Theorem~4 they count the cubes of one level that meet a thin cylinder around an edge. One representative per cube then bounds the nearby spanner edges of that level.

\item \textbf{Result transfer: Not established.} Theorem~4 relies on a hierarchy whose cubes at one level all have the same size, with one fixed representative per cube.

\item \textbf{Compressed definition: Uncompressed hierarchy.} In Section~3.1 they review a quadtree with cubes of side $a_0/2^\ell$ at level~$\ell$ and one representative per non-empty cube.
\end{itemize}

\subparagraph{Online Spanners in Metric Spaces~\cite{Bhore2022Online}}
\begin{itemize}
\item \textbf{Model and assumptions.} The paper names no model of computation. The Euclidean algorithm uses a quadtree (Section~2.1), and for an efficient implementation they refer to Fischer and Har-Peled~\cite{FischerHarPeled2005} (Section~1.2.2).

\item \textbf{Stated result.} Theorem~2: the online algorithm maintains a $(1+O(\varepsilon))$-spanner of weight $O_d(\varepsilon^{-d}\log n)\cdot\|\mathrm{MST}\|$ with $O_d(\varepsilon^{1-d}\log\varepsilon^{-1})\cdot n$ edges.

\item \textbf{Usage of packing properties: Yes.} In the proof of Theorem~2 they pick representatives in pairwise non-neighbouring cells of one level and place disjoint balls around them, which bounds the weight of the minimum spanning tree from below.

\item \textbf{Result transfer: Not established.} Their analysis uses a hierarchy with fixed cells and representatives chosen online.

\item \textbf{Compressed definition: Uncompressed hierarchy.} They use cubes of side $a_0/2^\ell$ at level~$\ell$.
\end{itemize}

\subparagraph{Dynamic Planar Point Location with Sub-Logarithmic Local Updates~\cite{LofflerSimonsStrashFull}}\label{Loffler}
\begin{itemize}
\item \textbf{Model and assumptions.} They target a real-valued pointer machine with comparisons, addition, subtraction and halving of reals (Appendix~A). The main text uses compressed cells aligned by floor. In Appendix~C.1 they allow non-aligned components and align two components when they merge, at a cost of $O(1)$ per node of the smaller one.

\item \textbf{Stated result.} Theorem~4.7: for $n$ disjoint convex $\beta$-thick objects in the plane, a structure of size $O(\beta n)$ supports insertions, deletions and point location in $O(\beta\log n)$ time and $\rho$-similar updates in $O(\beta\log\log n+\log(\beta\rho))$ time, all in the worst case.

\item \textbf{Usage of packing properties: Yes.} In Lemmas~4.1 and~4.2 they relate the size of a region to the size of the cell storing it, and cover a region by $O(\beta)$ cells of that size. In Lemma~4.3 they bound the regions meeting a leaf by $O(\beta)$, using convexity and disjointness.

\item \textbf{Result transfer: Not established.} Our tree does not maintain the storing cells, the region tags or the marked-ancestor structures of Section~4.

\item \textbf{Compressed definition: Compressed child at any scale.} In Appendix~A they propose the interval spanned by the smallest and largest element. In Appendix~C.1 they allow any sufficiently small square inside the parent. The alignment of merged components is described by its cost.
\end{itemize}

\subsection{Smallest bounding squares}
\label{app:literature-lowerbound}
L\"offler, Simons and Strash~\cite[Appendix~A]{LofflerSimonsStrashFull} propose, for a model without floor, the interval spanned by the smallest and largest element of a point set as its cell.
Theorem~\ref{thm:strict-lower-bound} applies when this choice is maintained together with all requirements of Definition~\ref{def:strict}.
In Appendix~C.1 they also allow any sufficiently small square inside the parent, which we discuss this \hyperref[Loffler]{subparagraph} in Appendix~\ref{app:literature-gaps}.

\end{document}